\documentclass[]{aa}

\usepackage{amsmath}
\usepackage[dvips]{graphicx}
\usepackage{natbib}
\bibpunct{(}{)}{;}{a}{}{,}
\usepackage[american]{babel}
\usepackage{txfonts}

\usepackage{setspace}\usepackage{threeparttable}

\def\lapprox{\lower.4ex\hbox{$\;\buildrel <\over{\scriptstyle\sim}\;$}}
\def\gapprox{\lower.4ex\hbox{$\;\buildrel >\over{\scriptstyle\sim}\;$}}
\def\sigmapar{\sigma_{||}}
\def\sigmaperp{\sigma_\perp}
\def\sigmaT{\sigma_{\rm T}}

\def\tauperp{\tau_\perp}

\def\vff{v_{\rm ff}}
\def\LX{L_{\rm X}}
\def\Lcrit{L_{\rm crit}}
\def\Lcoul{L_{\rm coul}}
\def\Ecyc{E_{\rm cyc}}
\def\veff{v_{\rm eff}}
\def\hc{h_{\rm c}}
\def\hs{h_{\rm s}}
\def\Pr{P_{\rm r}}
\def\mp{m_{\rm p}}
\def\vps{v_{\rm ps}}
\def\tesc{t_{\rm esc}}

\begin{document}

\title{Spectral formation in accreting X-ray pulsars: bimodal variation
of the cyclotron energy with luminosity}

\titlerunning{Bimodal spectral variation in X-ray pulsars}

\author{P. A. Becker\inst{1}
	\and
        D. Klochkov\inst{2}
        \and
        G. Sch\"onherr
        \inst{3}
        \and
        O. Nishimura
        \inst{4}
        \and
        C. Ferrigno
        \inst{5}
        \and
        I. Caballero
        \inst{6}
        \and
        P. Kretschmar
        \inst{7}
        \and
        M. T. Wolff
        \inst{8}
        \and
        J. Wilms
        \inst{9}
        \and
        R. Staubert\inst{2}
      }

\authorrunning{P. A. Becker et al.}
  \offprints{P. A. Becker}

\institute{School of Physics, Astronomy, and Computational Sciences,
MS 5C3, George Mason University, \\
4400 University Drive, Fairfax, VA, USA \\
	\email{pbecker@gmu.edu}
	\and
Institut f\"ur Astronomie und Astrophysik, Abt. Astronomie, Universit\"at
T\"ubingen, Sand 1, 72076 T\"ubingen, Germany
	\and
Leibniz-Institut f\"ur Astrophysik Potsdam, An der Sternwarte 16,
14482 Potsdam, Germany
	\and
Department of Electronics and Computer Science, Nagano National College
of Technology, 716 Tokuma, Nagano 381-8550, Japan
	\and
ISDC Data Center for Astrophysics, Universit\'e de Gen\`eve, Chemin d'Ecogia
16, 1290 Versoix, Switzerland
	\and
AIM (UMR 7158 CEA/DSM - CNRS - Universit´e Paris Diderot) Irfu/Service
d'Astrophysique, F-91191 Gif-sur-Yvette, France
	\and
European Space Agency, European Space Astronomy Centre, P.O. Box 78,
28691 Villanueva de la Ca\~nada, Madrid, Spain
	\and
Space Science Division, Naval Research Laboratory,
Washington, DC, USA
	\and
Dr. Karl Remeis-Observatory and Erlangen Centre for Astroparticle
Physics, Sternwartstr. 7, 96049 Bamberg, Germany
          }

\date{Received ---; accepted ---}

\abstract
{Accretion-powered X-ray pulsars exhibit significant variability of the
Cyclotron Resonance Scattering Feature (CRSF) centroid energy on
pulse-to-pulse timescales, and also on much longer timescales. Two types
of spectral variability are observed. For sources in group 1, the CRSF
energy is negatively correlated with the variable source luminosity, and
for sources in group 2, the opposite behavior is observed. The physical
basis for this bimodal behavior is currently not well understood.}
{We explore the hypothesis that the accretion dynamics in the group 1
sources is dominated by radiation pressure near the stellar surface, and
that Coulomb interactions decelerate the gas to rest in the group 2
sources.}
{We derive a new expression for the critical luminosity, $\Lcrit$, such
that radiation pressure decelerates the matter to rest in sources with
X-ray luminosity $\LX > \Lcrit$. The formula for $\Lcrit$ is based on a
simple physical model for the structure of the accretion column in
luminous X-ray pulsars that takes into account radiative deceleration,
the energy dependence of the cyclotron cross section, the thermodynamics
of the accreting gas, the dipole structure of the pulsar magnetosphere,
and the diffusive escape of radiation through the column walls. We show
that for typical neutron star parameters, $\Lcrit = 1.5 \times 10^{37}
B_{12}^{16/15}\,\rm erg\ sec^{-1}$, where $B_{12}$ is the surface
magnetic field strength in units of $10^{12}\,$G.}
{The formula for the critical luminosity is evaluated for 5 sources,
using the maximum value of the CRSF centroid energy to estimate the
surface magnetic field strength $B_{12}$. The results confirm that the
group 1 sources are supercritical ($\LX > \Lcrit$) and the group 2
sources are subcritical ($\LX < \Lcrit$), although the situation is less
clear for those highly variable sources that cross over the line $\LX =
\Lcrit$. We also explain the variation of the CRSF energy as a
consequence of variation of the characteristic emission height with
luminosity. The sign of this dependence is opposite in the supercritical
and subcritical cases, hence creating the observed bimodal behavior.}
{We have developed a new model for the critical luminosity in
accretion-powered X-ray pulsars that explains the bimodal dependence of
the CRSF centroid energy on the X-ray luminosity $\LX$. Our model
provides a physical basis for the observed variation of the CRSF energy
as a function of $\LX$ for both the group 1 (supercritical) and the
group 2 (subcritical) sources as a result of variation of the
characteristic emission height in the column.}

\keywords{X-rays: binaries, pulsars: individual: Her X-1, V\,0332+53,
A\,0535+26, 4\,U0115+63, GX\,304-1}

\maketitle

\section{Introduction}

\label{sec:intro}
X-ray binary pulsars (XRBPs) were first observed by Giacconi et al.
(1971) and Tananbaum et al. (1972), and now include many of the
brightest sources in the X-ray sky. In XRBPs, the main sequence
companion star transfers matter to the neutron star via Roche lobe
overflow, or via a strong stellar wind (Frank et al. 2002). The gas
forms an accretion disk around the neutron star, and the material
spirals inward until the pressure of the star's dipole magnetic field
becomes comparable to the ram pressure of the matter in the disk. This
occurs at the Alfv\'en radius, located several thousand kilometers out
in the accretion disk. The fully-ionized accreting plasma is entrained
by the magnetic field at the Alfv\'en radius, and from there the matter
is guided through the magnetosphere, forming accretion columns at one or
both of the magnetic poles of the star. As the star spins, the
inclination angle between the star's magnetic axis and the axis of the
accretion disk changes, and therefore the Alfv\'en radius varies with
the spin period of the star.

The observed X-ray emission is powered by the conversion of
gravitational potential energy into kinetic energy, which is then
transferred to the radiation field via electron scattering, and
ultimately escapes through the walls of the column. The structure of the
accretion column is maintained by the strong magnetic field, with a
surface strength $B_* \gapprox 10^{12}\,$G, which results in a magnetic
pressure far exceeding that of either the gas or the radiation field.
The high incident speed of the freely-falling plasma, $\sim 0.6\,c$,
creates very high temperatures, $T \sim 10^8\,$K. However, the observed
X-ray pulsar spectra are highly nonthermal, indicating that the
accreting gas is unable to equilibrate during the accretion timescale.
In this situation, bulk and thermal Comptonization play key roles in
establishing the shape of the observed spectra (Becker \& Wolff 2007).

The X-ray spectra of many XRBPs contain \emph{cyclotron resonant
scattering features} (CRSFs) appearing as absorption lines. The features
are caused by resonant scattering of photons off plasma electrons whose
energy is quantized according to their Landau level (see e.g. Tr\"umper
et al. 1978; Isenberg et al. 1998; Araya-G\'ochez \& Harding 2000). The
CRSFs, when detected, provide a direct measurement of the magnetic field
strength at the characteristic altitude of the X-ray emission. The
energy of the fundamental line and the spacing between the harmonics are
approximately proportional to the $B$-field strength.

Many XRBPs display X-ray spectra that vary significantly with luminosity
on timescales much longer than the pulsation period. In particular,
variations in the energy of CRSFs as a function of luminosity on
timescales of days to months have been detected in V\,0332+53 (Mowlavi
et al. 2006; Tsygankov et al. 2010), 4U\,0115+63 (Mihara et al. 2004;
Tsygankov et al. 2007), and Her X-1 (Staubert et al. 2007; Vasco et al.
2011). In addition to the longer-term variability, there is also
mounting evidence for pulse-to-pulse variability, in which the spectral
hardness, the centroid energy of the CRSF, and the luminosity vary in a
correlated way (Klochkov et al. 2011). This short-timescale variability
is likely related to the non-stationarity of the accretion flow, perhaps
indicating that the entrainment of matter from the disk onto the
magnetic field lines results in filaments and blobs of accreting gas
which are then channeled onto the star in a non-uniform way.

The data from both long-term and short-term (pulse-to-pulse)
observations point to the existence of two types/modes of spectral
variability (see discussion by Klochkov et al. 2011). For sources in
group 1 (e.g., V\,0332+53), the centroid energy of the CRSF is
negatively correlated with luminosity. For sources in group 2 (e.g.,
Her\,X-1), the opposite behavior is observed. The type of spectral
variability is likely driven by the mode of accretion, which in turn is
determined by the luminosity (see discussion in Staubert et al. 2007).
Staubert et al. (2007) and Klochkov et al. (2011) have proposed that for
sources in group~1, the deceleration of the flow to rest at the stellar
surface is accomplished by the pressure of the radiation field, and in
the group~2 sources the deceleration occurs via Coulomb interactions. In
this interpretation, a given source falls in one group or the other
depending on the value of its X-ray luminosity, $\LX$, relative to the
critical luminosity, $\Lcrit$. The hypothesis is that the group~1
sources are supercritical, with X-ray luminosity $\LX > \Lcrit$, and the
group~2 sources are subcritical ($\LX < \Lcrit$).

The theory predicts that sources in their supercritical state should
display a negative correlation between the luminosity and the cyclotron
energy, while sources in the subcritical state should display the
reverse behavior. Geometrically, the variation of the CRSF energy with
luminosity is connected with variation of the characteristic emission
height, which is the altitude in the accretion column where the
cyclotron absorption feature is imprinted on the observed spectrum. The
variation of the emission height as a function of luminosity in the
subcritical and supercritical cases is indicated schematically in
Fig.~1.

The general picture described above provides a qualitative basis for the
interpretation of the observed correlated variation of the CRSF centroid
energy with X-ray luminosity in some XRBPs. However, in order to obtain
a quantitative understanding of these observations, one must develop a
more detailed physical model for the critical luminosity, and for the
dependence of the CRSF energy on the luminosity in the subcritical and
supercritical regimes. The first goal of this paper is to derive a new
expression for the critical luminosity, taking into account the
magnetospheric connection between the radius of the accretion column and
the Alfv\'en radius in the disk, and the energy and angle dependence of
the cyclotron scattering cross section. The second goal is to examine
the dependence of the CRSF centroid energy on the luminosity in the
subcritical and supercritical sources.

The remainder of the paper is organized as follows. In Sect.~2, we
obtain a fundamental expression for the critical luminosity that depends
on the stellar mass, radius, and surface magnetic field strength. In
Sect.~3 we develop simple physical models for the variation of the
characteristic emission height as a function of the luminosity for
subcritical and supercritical sources. In Sect.~4, we evaluate the
critical luminosity based on measurements of the CRSF centroid energy
for several XRBPs. We use our subcritical and supercritical models for
the variation of the emission height to predict the variation of the
CRSF energy as a function of luminosity. The predicted spectral
variability is compared with the observational data for each source. We
discuss our results and draw conclusions in Sect.~5.

\section{Critical luminosity}
\label{model}

\subsection{Eddington luminosity and radiative deceleration}

First we recall the definition of the standard Eddington luminosity,
$L_{\rm Edd}$, for spherically symmetric accretion onto a central mass
$M_*$. If the accreting gas is fully-ionized hydrogen, we obtain
\begin{equation}
L_{\rm Edd} = {4 \pi G M_* \mp c \over \sigmaT}
\ ,
\label{eq1}
\end{equation}
where $\sigmaT$ is the Thomson cross section, $\mp$ is the proton mass,
$c$ is the speed of light, and $G$ is the gravitational constant. When
the X-ray luminosity $\LX=L_{\rm Edd}$, the rate at which momentum is
transferred to the gas via Compton scattering balances the gravitational
force. Hence if $\LX>L_{\rm Edd}$, the net force is in the outward
direction and the gas decelerates as it falls toward the central mass.

We need to make two adjustments to Equation~(\ref{eq1}) in order to
compute the effective Eddington limit, $L^{*}_{\rm Edd}$, appropriate
for treating X-ray pulsar accretion columns. The first adjustment is to
replace the Thomson cross section $\sigmaT$ with $\sigmapar$, which
represents the mean scattering cross section for photons propagating
parallel to the magnetic field axis. The second adjustment is to take
the geometry of the accretion flow into account by reducing the
luminosity by the ratio of the column cross-sectional area divided by
the surface area of the star. Employing these corrections yields for the
effective Eddington limit
\begin{equation}
L^{*}_{\rm Edd} = L_{\rm Edd} \, {\sigmaT \over \sigmapar}
\, {\pi r_0^2 \over 4 \pi R_*^2}
= {G M_* \mp c \over \sigmapar} \, {\pi r_0^2
\over R_*^2} \ ,
\label{eq2}
\end{equation}
where $R_*$ is the stellar radius and $r_0$ denotes the radius of the
accretion column, which we assume to have an approximately cylindrical
geometry.

Based on Equation~(\ref{eq2}), Basko \& Sunyaev (1976) concluded that
for X-ray luminosities $\LX \gapprox 10^{36}\rm erg\ s^{-1}$, the
incident, freely-falling gas is decelerated by a vertical flux of
radiation that is locally super-Eddington. The scattering of the
incident radiation removes kinetic energy from the electrons (and from
the protons via Coulomb coupling), thereby decelerating the gas.
Although the radiation flux seen by the gas inside the column is
propagating upward, the X-rays that ultimately carry away the kinetic
energy actually escape through the walls of the column, rather than the
top, unless the luminosity $\LX \lapprox 10^{36}\rm erg\ s^{-1}$.

The deceleration of the gas begins when the freely-falling material
encounters a radiation-dominated shock whose height above the star
increases with increasing luminosity, reaching an altitude of several
kilometers for $\LX \sim 10^{37-38}\rm erg\ s^{-1}$. As the gas passes
through the shock, the accretion velocity is reduced by a factor of
$\sim 7$. Even though the radiation flux inside the column is
super-Eddington, the height of the radiation-dominated shock is stable.
This reflects the fact that the shock is a wave structure, and is not
composed of a fixed population of particles. Matter moves through the
shock and decelerates, but the shock height remains fixed, unless the
luminosity changes. Unlike a classical gas-mediated shock, the
radiation-dominated shock is not discontinuous, and instead has a
thickness that is a few times larger than the mean-free path for
radiation scattering.

\begin{figure*}
\begin{center}$
\begin{array}{cc}
\includegraphics[width=2.5in]{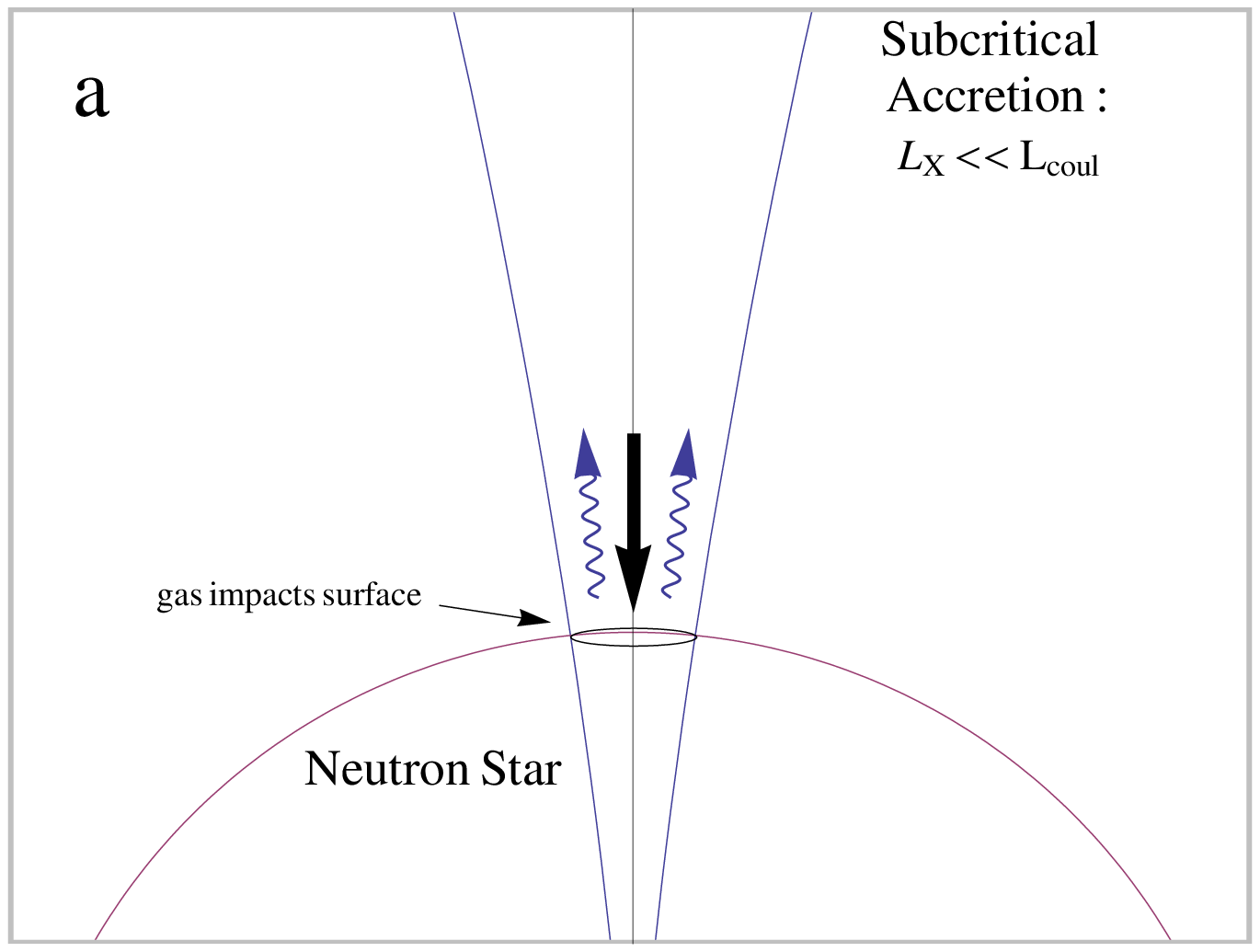} &
\includegraphics[width=2.5in]{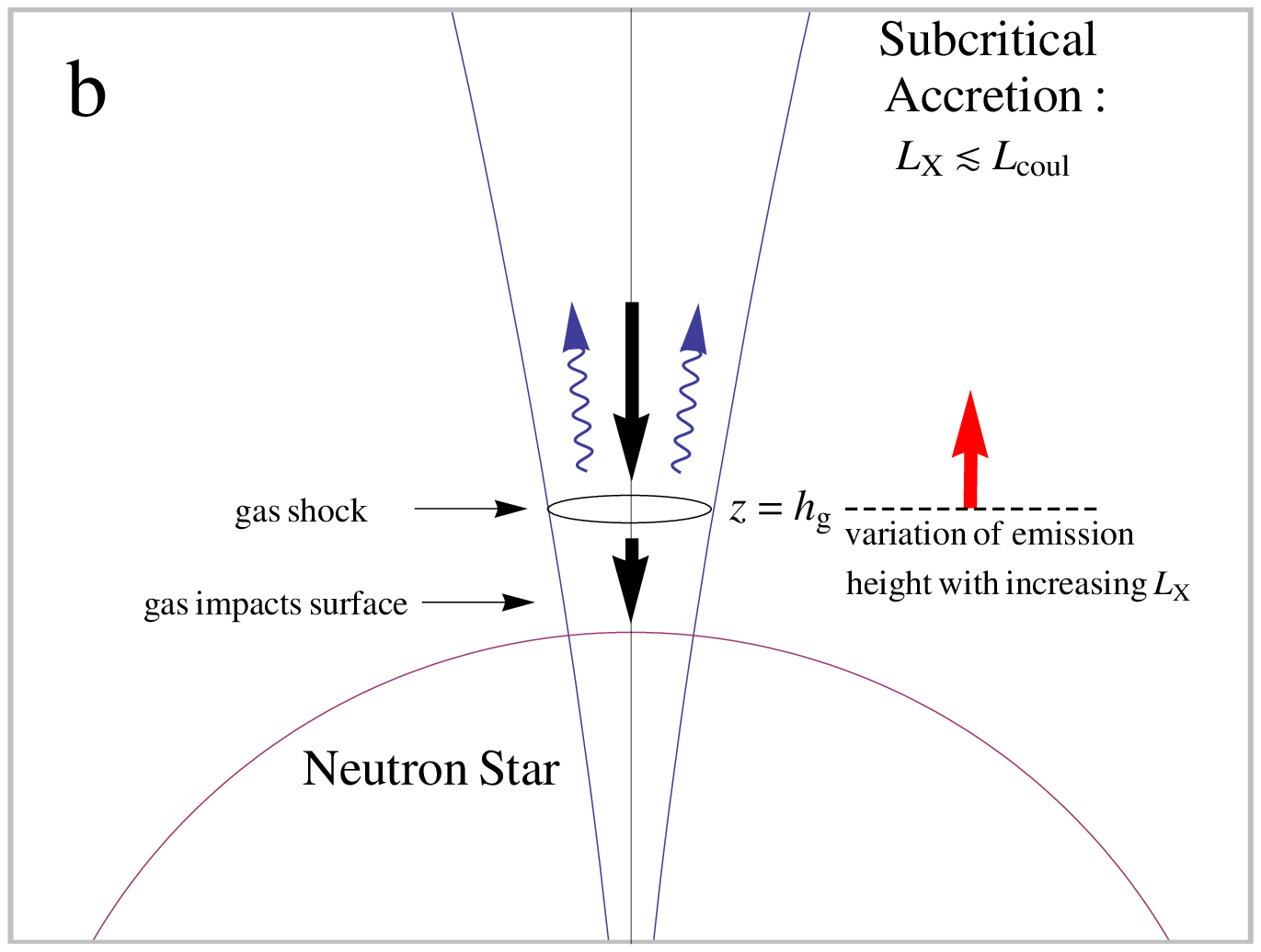} \\
\includegraphics[width=2.5in]{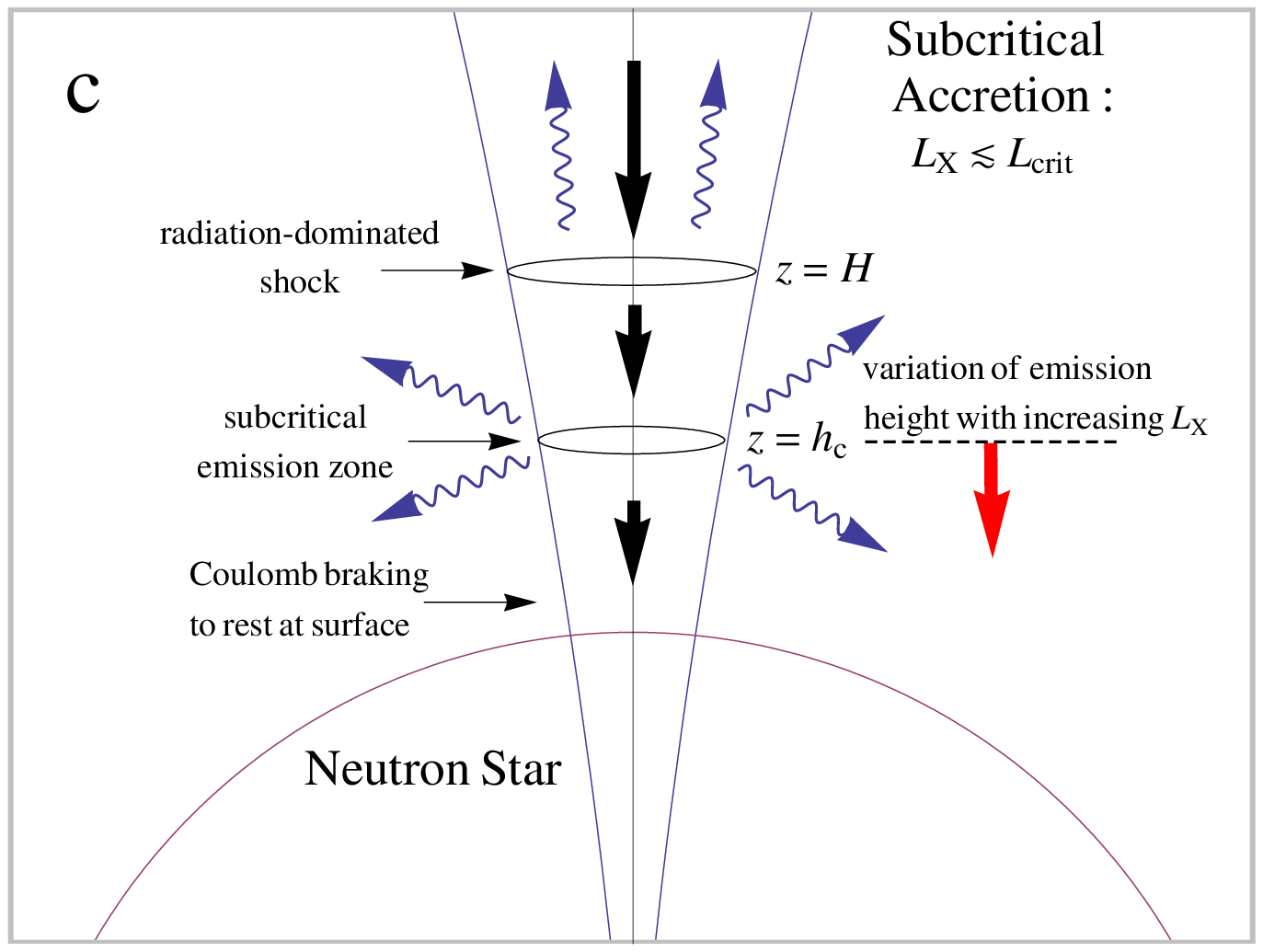} &
\includegraphics[width=2.5in]{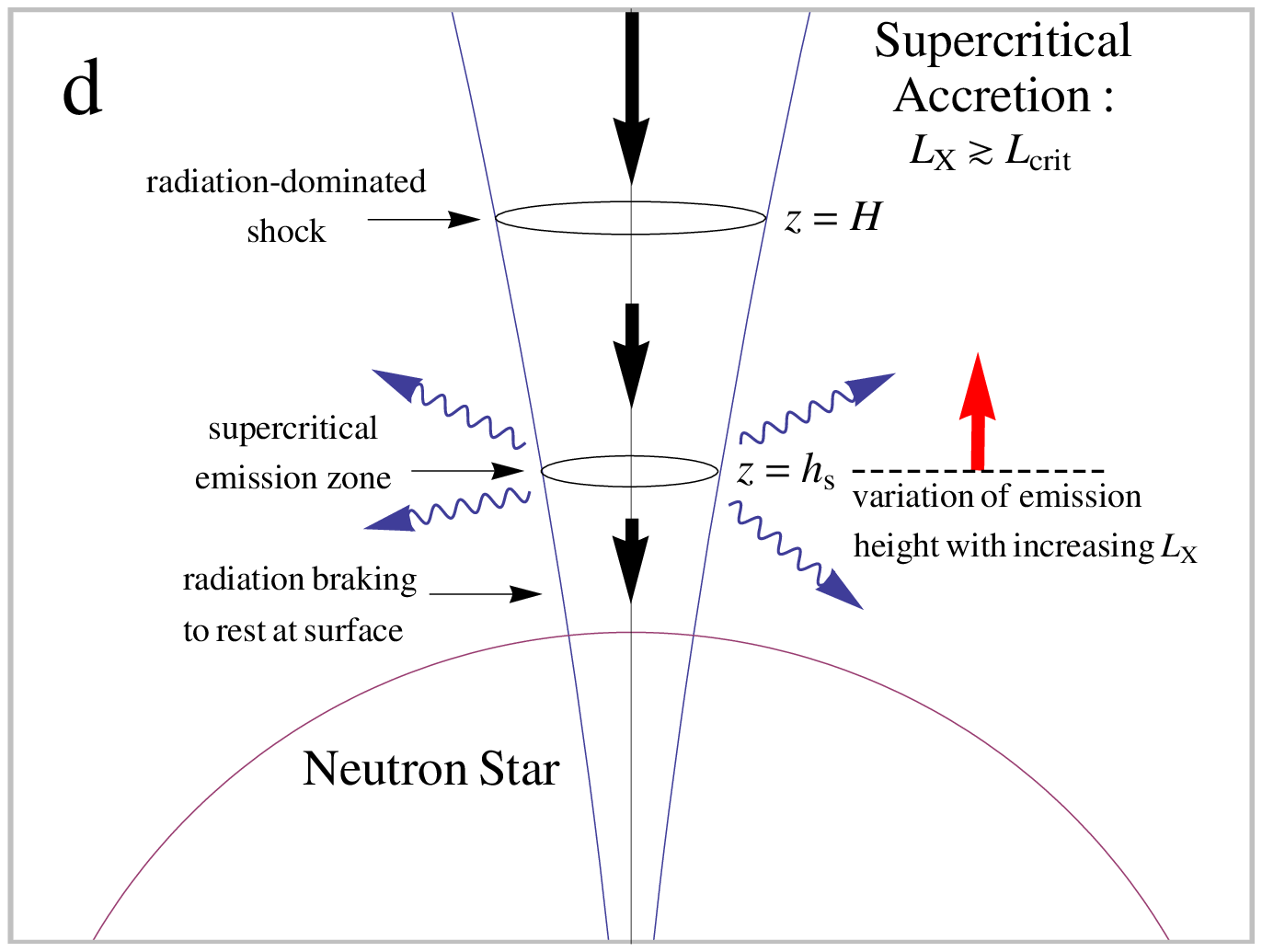}
\end{array}$
\label{fig:diagrams}
\end{center}
\caption{Schematic illustration of the geometry of the accretion column
and the variation of the characteristic emission height and emission
beam pattern with increasing luminosity $\LX$: (a) subcritical, $\LX <
\Lcoul < \Lcrit$, pencil beam; (b) subcritical, $\LX \lapprox \Lcoul <
\Lcrit$, pencil beam; (c) subcritical, $\Lcoul < \LX \lapprox \Lcrit$,
intermediate beam pattern; (d) supercritical, $\LX \gapprox \Lcrit$, fan
beam.}
\end{figure*}

Below the radiation-dominated shock, the matter is further decelerated
in the hydrostatic ``sinking regime,'' in which the remaining momentum
is transferred to the radiation field and radiated away through the
column walls (Basko \& Sunyaev 1976). The specific mechanism
accomplishing the final deceleration to rest at the stellar surface in
the sinking region depends on the luminosity of the accretion flow (see
Fig.~1). At the highest luminosities, $\LX \sim 10^{37-38}\rm erg\
sec^{-1}$, the radiation field accomplishes the deceleration all the way
down to the stellar surface (Basko \& Sunyaev 1976). At intermediate
luminosities $\LX \sim 10^{36-37}\rm erg\ sec^{-1}$, the final phase of
deceleration may occur via Coulomb breaking in a plasma cloud just above
the stellar surface (Nelson et al. 1993). It is expected that at very
low luminosities, $\LX \lapprox 10^{34-35}\rm erg\ sec^{-1}$, there is
no radiation-dominated shock at all, and the material passes through a
conventional gas-mediated shock at altitude $z=h_g$ before striking the
stellar surface (Langer \& Rappaport 1982).

The angular pattern of the emitted radiation also depends on the
luminosity (see Fig.~1). In high-luminosity sources ($\LX \sim 10^{37-38}\rm erg\
s^{-1}$), the emitted radiation primarily escapes through the column
walls in the sinking region, forming a ``fan beam'' (Davidson 1973). For
low-luminosity sources ($\LX \lapprox 10^{35}\rm erg\ s^{-1}$), the
emission escapes from the top of the column, forming a ``pencil beam''
(Burnard et al. 1991; Nelson et al. 1993). In the intermediate range,
$\LX \lapprox 10^{35-37}\rm erg\ s^{-1}$, the emission pattern may be a
hybrid combination of these two types (Blum \& Kraus 2000).

Focusing on the high-luminosity case for now, we can estimate the
luminosity required to decelerate the gas to rest at the stellar surface
by considering the physical processes occurring in the sinking region
below the radiation-dominated shock. The accreting matter approaches the
top of the shock with the incident free-fall velocity, which we approximate
using the value at the stellar surface,
\begin{equation}
\vff = \left(2 G M_* \over R_*\right)^{1/2}
\ .
\label{eq3}
\end{equation}
Advection is dominant over diffusion in the shock, and therefore very
little radiation energy escapes through the walls of the accretion
column in the vicinity of the shock (Burnard et al. 1991). Hence the
shock jump conditions are well approximated by the standard
Rankine-Hugoniot relations for a gas with adiabatic index $\gamma=4/3$
(Basko \& Sunyaev 1976). In this case, the matter leaves the shock with
the post-shock velocity
\begin{equation}
\vps = {1 \over 7} \, \vff
= {1 \over 7} \left(2 G M_* \over R_*\right)^{1/2}
\ ,
\label{eq4}
\end{equation}
where we have assumed that the shock is strong, which is reasonable in
the luminous sources (Becker 1998). If the altitude of the
radiation-dominated shock above the stellar surface is $H$, and the gas
decelerates at a constant rate $a$ from the post-shock velocity $\vff/7$
to rest at the stellar surface in the dynamical time $t_{\rm dyn}$, then
we can write the simple kinematical relations
\begin{equation}
H = {1 \over 2} \, a \, t_{\rm dyn}^2 \ , \ \ \ \ \ 
\vps = a \, t_{\rm dyn} \ .
\label{eq5}
\end{equation}
Upon elimination of $t_{\rm dyn}$, we obtain for the required upward
acceleration
\begin{equation}
a = {\vps^2 \over 2 H} = {G M_* \over 49 R_* H}
\ .
\label{eq6}
\end{equation}

Since the effective gravity is reduced by the pressure of the radiation
field, the net acceleration can also be related to the luminosity
$\LX$ via
\begin{equation}
a = \left({\LX \over L^{*}_{\rm Edd}} - 1\right) {G M_* \over R_*^2}
\ .
\label{eq7}
\end{equation}
Setting Eqs.~(\ref{eq6}) and (\ref{eq7}) equal and solving for $\LX$
yields
\begin{equation}
\LX = \Lcrit \equiv L^{*}_{\rm Edd} \left({R_* \over 49 H}
+ 1\right)
\ .
\label{eq8}
\end{equation}
Substituting for $L^{*}_{\rm Edd}$ using Eq.~(\ref{eq2}), we obtain
for the critical luminosity
\begin{equation}
\Lcrit = {G M_* \mp c \over \sigmapar}
\, {\pi r_0^2 \over R_*^2} \left({R_* \over 49 H} + 1\right)
\ .
\label{eq9}
\end{equation}
Our goal is to express the parameters $r_0$, $\sigmapar$, and $H$
appearing on the right-hand side of Eq.~(\ref{eq9}) in terms of
observable quantities.

\subsection{Radiation-dominated shock height}

The altitude, $H$, of the radiation-dominated shock can be estimated by
considering the relationship between the dynamical timescale for
deceleration, $t_{\rm dyn}$, and the photon escape timescale, $t_{\rm
diff}$, which is the mean time it takes the photons to diffuse through
the walls of the accretion column. In order for the gas to come to rest
at the stellar surface, these two timescales must be comparable in the
sinking region below the shock, which is a general property of accretion
flows onto white dwarf stars and neutron stars (e.g., Imamura et al.
1987). Combining Eqs.~(\ref{eq4}) and (\ref{eq5}), we obtain for
the dynamical time
\begin{equation}
t_{\rm dyn} = {2 H \over \vps}
= 14 H \left(R_* \over 2 G M_*\right)^{1/2}
\ .
\label{eq10}
\end{equation}
The escape timescale for the photons to diffuse through the column walls
is estimated by writing
\begin{equation}
\tesc = {r_0 \over v_\perp^{\rm diff}} \ , \ \ \ \ \
v_\perp^{\rm diff} = {c \over \tauperp} \ , \ \ \ \ \
\tauperp = r_0 n_e \sigmaT \ ,
\label{eq11}
\end{equation}
where $n_e$ is the electron number density, $v_\perp^{\rm diff}$ is the
photon diffusion velocity perpendicular to the column axis, and
$\tauperp$ is the perpendicular optical thickness. The Thomson cross
section $\sigmaT$ is appropriate for photons propagating perpendicular
to the column axis (Wang \& Frank 1981). The electron number density
$n_e$ appearing in Eq.~(\ref{eq11}) can be eliminated using the mass
conservation relation,
\begin{equation}
\dot M = \pi r_0^2 n_e \mp v
\ ,
\label{eq12}
\end{equation}
where $\dot M$ is the accretion rate and $v$ is the inflow velocity,
defined to be positive. Combining relations, we can express the escape
time through the walls as
\begin{equation}
\tesc = {\dot M \sigmaT \over \pi \mp v c}
\ .
\label{eq13}
\end{equation}

The deceleration in the sinking region begins on the downstream side of
the shock, and therefore we set $v=\vps$ in Eq.~(\ref{eq13}) and
equate $t_{\rm dyn}$ and $\tesc$ to obtain
\begin{equation}
H = {\dot M \sigmaT \over 2 \pi \mp c}
\ ,
\label{eq14}
\end{equation}
which is essentially the same result obtained by Burnard et al. (1991).
Expressing the accretion rate in terms of the luminosity using the
relation
\begin{equation}
\LX = {G M_* \dot M \over R_*}
\label{eq15}
\end{equation}
yields the equivalent expression
\begin{equation}
H = 1.14 \times 10^5 \, {\rm cm} \,
\left(M_* \over 1.4 M_\odot\right)^{-1}
\left(R_* \over 10\,{\rm km}\right)
\left(\LX \over 10^{37} {\rm erg\ sec^{-1}}\right)
\ .
\label{eq16}
\end{equation}
This confirms that the shock is located a few kilometers above the
stellar surface in the luminous sources with $\LX \sim 10^{37-38}\rm erg\
s^{-1}$ (Basko \& Sunyaev 1976). It follows that $R_* /(49 H) \ll 1$ for
sources close to or above the critical luminosity, and therefore
Eq.~(\ref{eq9}) reduces to
\begin{equation}
\Lcrit = {G M_* \mp c \over \sigmapar}
\, {\pi r_0^2 \over R_*^2}
\ ,
\label{eq17}
\end{equation}
in agreement with Burnard et al. (1991). Note that in this limit, the
critical luminosity simply reduces to the effective Eddington value given by
Eq.~(\ref{eq2}).

\subsection{Connection between column radius and Alfv\'en radius}

In this section, we wish to relate the critical luminosity $L_{\rm
crit}$ in Eq.~(\ref{eq17}) to the magnetic field strength at the stellar
surface, $B_*$, by utilizing the connection between the radius of the
accretion column, $r_0$, and the Alfv\'en radius in the disk, $R_{\rm
A}$. The inclination angle between the axis of the accretion disk and
the star's magnetic axis varies with a period equal to the pulsar's
spin period, and this causes an associated variation of the Alfv\'en
radius. However, for our purposes here, an adequate approximation is
obtained by using Eq.~(13) from Lamb et al. (1973), which yields
\begin{equation}
\begin{aligned}
R_{\rm A} = 2.73 \times 10^7 \ {\rm cm} \ \left(\Lambda \over 0.1\right)
\left(M_* \over 1.4 M_\odot \right)^{1/7}
& \left(R_* \over 10\,{\rm km}\right)^{10/7} \\
\times \ \left(B_* \over 10^{12}{\rm G} \right)^{4/7}
& \left(\LX \over 10^{37} {\rm erg\ sec^{-1}}\right)^{-2/7}
\ ,
\end{aligned}
\label{eq18}
\end{equation}
where the constant $\Lambda=1$ for spherical accretion and $\Lambda < 1$
for disk accretion. A variety of uncertainties are folded into
$\Lambda$, such as the spin-averaging of $R_{\rm A}$ and the possible
role of plasma shielding and other magnetospheric effects. Based on
Eq.~(2) from Harding et al. (1984), $\Lambda$ can be approximated in the
disk application using
\begin{equation}
\Lambda \approx 0.22 \ \alpha^{18/69}
\ ,
\label{eq19}
\end{equation}
where $\alpha < 1$ denotes the Shakura-Sunyaev parameter (Shakura \&
Sunyaev 1973). Although it is difficult to estimate $\alpha$ with any
certainty, we generally expect to find $\alpha \sim 0.01-0.1$. We
therefore set $\Lambda=0.1$ in our numerical applications.

The Alfv\'en radius in the disk is connected with the outer surface of
the accretion column through the dipole shape of the pulsar magnetosphere.
The equation for the shape of the critical field line as a function of
the polar angle $\theta$ is given by the standard dipole formula
\begin{equation}
R = R_{\rm A} \sin^2\theta
\ .
\label{eq20}
\end{equation}
Setting the radius $R$ equal to the stellar radius $R_*$ yields for the
critical angle at the outer edge of the accretion column
\begin{equation}
\sin^2\theta_c = {R_* \over R_{\rm A}}  
\ .
\label{eq21}
\end{equation}
Using the small-angle relation $\theta_c \approx \sin\theta_c$, we
obtain for the column radius
\begin{equation}
r_0 = R_* \, \theta_c = R_* \left(R_* \over R_{\rm A}\right)^{1/2}
\ .
\label{eq22}
\end{equation}

By substituting for the Alfv\'en radius in Eq.~(\ref{eq22}) using
Eq.~(\ref{eq18}), we find that the expression for the column radius
$r_0$ can be rewritten in cgs units as
\begin{equation}
\begin{aligned}
r_0 = 1.93 \times 10^5 \, {\rm cm} \ \left(\Lambda \over 0.1\right)^{-1/2}
\left(M_* \over 1.4 M_\odot\right)^{-1/14}
& \left(R_* \over 10\,{\rm km}\right)^{11/14} \\
\times \ \left(B_* \over 10^{12}{\rm G}\right)^{-2/7}
& \left(\LX \over 10^{37} {\rm erg\ sec^{-1}}\right)^{1/7}
\ .
\end{aligned}
\label{eq23}
\end{equation}
Using Eq.~(\ref{eq23}) to substitute for $r_0$ in Eq.~(\ref{eq17}), and
setting $\LX=\Lcrit$, we obtain for the critical luminosity the new
expression
\begin{equation}
\begin{aligned}
\Lcrit = 7.79 \times 10^{35} \, {\rm erg\ sec^{-1}}
\left(\Lambda \over 0.1\right)^{-7/5}
& \left({\sigmapar \over \sigmaT}\right)^{-7/5} \\
\times \ \left(M_* \over 1.4 M_\odot\right)^{6/5}
& \left(R_* \over 10\,{\rm km}\right)^{-3/5}
\left(B_* \over 10^{12}{\rm G}\right)^{-4/5}
\ ,
\end{aligned}
\label{eq24}
\end{equation}
where we have also introduced the Thomson cross section $\sigmaT$ as a
convenient scaling for the parallel scattering cross section
$\sigmapar$. The next step is to evaluate the cross section ratio
$\sigmapar/\sigmaT$ in terms of observable source parameters.

\subsection{Electron scattering cross section for parallel propagation}

In typical X-ray pulsars, most of the observed radiation is emitted at
energies below the cyclotron energy, $\Ecyc$. Hence the cross-section
ratio $\sigmapar / \sigmaT$ can be roughly approximated using (e.g.,
Arons et al. 1987)
\begin{equation}
{\sigmapar \over \sigmaT} = \left({\bar E \over \Ecyc}
\right)^2
\ ,
\label{eq25}
\end{equation}
where $\bar E$ is a measure of the mean energy of the photons
propagating parallel to the magnetic field, and the cyclotron energy
$\Ecyc$ is given by
\begin{equation}
\Ecyc = 11.58 \ {\rm keV} \left(B \over 10^{12} \, {\rm G}\right)
\ .
\label{eq26}
\end{equation}

The mean photon energy $\bar E$ in Eq.~(\ref{eq25}) can be estimated
observationally by integrating the spectrum for a given source. However,
in luminous X-ray pulsars, most of the observed radiation escapes
through the walls of the accretion column, perpendicular to the magnetic
field, and therefore the observed spectrum may not be representative of
the distribution of photons propagating along the column axis. As an
alternative, we can estimate $\bar E$ based on the thermal structure of
the accreting gas. Specifically, we assume that
\begin{equation}
\bar E = w k T_{\rm eff}
\ ,
\label{eq27}
\end{equation}
where $T_{\rm eff}$ is the effective temperature of the radiation in the
post-shock region, $k$ is Boltzmann's constant, and the constant $w$
depends on the shape of the spectrum inside the column. We expect that
$w \sim 1-3$, with the lower value corresponding to bremsstrahlung and
the upper value to a Planck spectrum. Detailed models suggest that the
spectrum inside the column is dominated by bremsstrahlung emission
(Becker \& Wolff 2007), and therefore we will set $w=1$ in the numerical
results presented later.

The effective temperature is related to the post-shock radiation
pressure, $\Pr$, via
\begin{equation}
a T_{\rm eff}^4 =  3 \Pr
\ .
\label{eq28}
\end{equation}
The value of $\Pr$ can be estimated using the momentum balance relation
\begin{equation}
\Pr = \rho_{\rm ff} \vff^2 = {\dot M \vff \over \pi r_0^2}
\ ,
\label{eq29}
\end{equation}
where $\rho_{\rm ff}$ and $\vff$ (Eq.~(\ref{eq3})) denote the upstream
mass density and velocity, respectively, just above the shock.
Eliminating $\Pr$ between Eqs.~(\ref{eq28}) and (\ref{eq29}), and
substituting for $\dot M$ and $r_0$ using Eqs.~(\ref{eq15}) and
(\ref{eq23}), we find that
\begin{equation}
\begin{aligned}
T_{\rm eff} = 4.35 \times 10^7 \ {\rm K} \
& \left(\Lambda
\over 0.1\right)^{1/4} \
\left(M_* \over 1.4 M_\odot\right)^{-5/56}
\left(R_* \over 10\,{\rm km}\right)^{-15/56} \\
& \times \ \left(B_* \over 10^{12}{\rm G}\right)^{1/7}
\left(\LX \over 10^{37} {\rm erg\ sec^{-1}}\right)^{5/28}
\ .
\end{aligned}
\label{eq30}
\end{equation}

Combining Eqs.~(\ref{eq25}), (\ref{eq26}), (\ref{eq27}), and
(\ref{eq30}), we obtain for the required cross section ratio
the result
\begin{equation}
\begin{aligned}
{\sigmapar \over \sigmaT} = 0.106 \ \left(\Lambda \over 0.1
\right)^{1/2} \
& w^2 \left(M_* \over 1.4 M_\odot\right)^{-5/28}
\left(R_* \over 10\,{\rm km}\right)^{-15/28} \\
& \times \ \left(B_* \over 10^{12}{\rm G}\right)^{-12/7}
\left(\LX \over 10^{37} {\rm erg\ sec^{-1}}\right)^{5/14}
\ .
\end{aligned}
\label{eq31}
\end{equation}
Using this result to substitute for $\sigmapar/\sigmaT$ in
Eq.~(\ref{eq24}) yields the final expression for the critical
luminosity as a function of the surface magnetic field strength,
\begin{equation}
\begin{aligned}
\Lcrit = 1.49 \times 10^{37} \, {\rm erg\ sec^{-1}}
& \left(\Lambda \over 0.1\right)^{-7/5} w^{-28/15} \\
\times \ \left(M_* \over 1.4 M_\odot\right)^{29/30}
& \left(R_* \over 10\,{\rm km}\right)^{1/10}
\left(B_* \over 10^{12}{\rm G}\right)^{16/15}
\ .
\end{aligned}
\label{eq32}
\end{equation}
For typical neutron star parameters, with $M_*=1.4\,M_\odot$,
$R_*=10\,$km, $\Lambda=0.1$, and $w=1$, we obtain $\Lcrit = 1.49 \times
10^{37} \, {\rm erg\ sec^{-1}} B_{12}^{16/15}$, where $B_{12}$ is the
surface magnetic field strength in units of $10^{12}\,$G. In Sect.~4 we
plot the critical luminosity and compare it with the variability data
for several XRBPs.

\section{Variation of emission height}

The new expression for the critical luminosity given by
Eq.~(\ref{eq32}) allows us to separate accretion-powered X-ray
pulsars into subcritical and supercritical categories. Our hypothesis is
that in the subcritical sources with variable luminosity $\LX$,
the cyclotron energy $\Ecyc$ will exhibit a positive correlation
with $\LX$, and in the supercritical sources the reverse behavior
will be observed. The observed CRSF is imprinted on the spectrum at the
altitude where most of the emitted radiation escapes from the accretion
column. In order to quantify the expected behaviors in the subcritical
and supercritical regimes, we must therefore explore the expected
variation of the emission height as a function of luminosity for both
types of sources. The geometry of the super- and subcritical sources is
depicted schematically in Fig.~1.

\subsection{Supercritical sources}

In the supercritical sources (luminosity $\LX \gapprox \Lcrit$),
radiation pressure dominates the flow dynamics all the way to the
stellar surface. Inside the radiation-dominated shock, the infalling
matter begins to decelerate as it first encounters the ``cushion'' of
radiation hovering at the shock altitude (Davidson 1973). At this
altitude, there is a local balance between downward advection and upward
diffusion of radiation, and therefore the photon distribution is roughly
static. Most of the kinetic energy of the accretion flow is carried away
by the scattered radiation, which is likely to be beamed in the downward
direction due to special-relativistic aberration (e.g., Ferrigno et al.
2009). Below the shock altitude, the photons are trapped by advection,
although they eventually manage to escape by diffusing through the walls
of the column. Hence the observed radiation does not escape from the
shock altitude $H$, but rather from a lower altitude.

Our goal here is to estimate the typical altitude, denoted by $\hs$, at
which the photons diffuse through the column walls to form the observed
X-ray spectrum in the supercritical case. We assume that the observed
CRSF is imprinted at this altitude, because at higher altitudes the
photons have not had enough time to diffuse through the column and
escape through the walls. Conversely, at lower altitudes, the increasing
density of the gas in the column inhibits the escape of radiation. We
therefore expect that the CRSF energy will reflect the cyclotron energy
at the altitude $z=\hs$ where the final deceleration phase begins.

We can estimate the emission height $\hs$ in the supercritical sources
by ensuring that all of the kinetic energy is radiated through the walls
by the escaping photons in the altitude range $0 < z < \hs$ (Basko \&
Sunyaev 1976). Working in the frame comoving with the plasma in the
accretion column, we note that the fraction of the radiation escaping
through the walls in the comoving time interval $dt'$ is equal to
$dt'/\tesc$, where $\tesc$ is the escape time given by Eq.~(\ref{eq13}).
The requirement that all of the radiation escapes by the time the matter
reaches the stellar surface is therefore expressed by the integral
condition
\begin{equation}
\int_0^{\hs} \left|{dt' \over dz'}\right|
{dz' \over \tesc(z')} = 1
\ .
\label{eq33}
\end{equation}
Using Eq.~(\ref{eq13}) to substitute for $\tesc$ yields
\begin{equation}
\int_0^{\hs} {\pi \mp v c \over \dot M \sigmaT}
{dz' \over \veff} = 1
\ ,
\label{eq34}
\end{equation}
where the effective velocity for the photon transport is defined by
\begin{equation}
\veff \equiv \left|{dz' \over dt'}\right|
\ .
\label{eq35}
\end{equation}

The flow is expected to be almost perfectly ``trapped'' in the
region below $z'=\hs$, meaning that advection and diffusion are nearly
balanced, leaving very little net flux of radiation (Becker 1998). This
implies that the effective velocity $\veff$ is much smaller than
the flow velocity $v$. We define the parameter $\xi$ as the ratio of
these two velocities,
\begin{equation}
\xi \equiv {\veff \over v}
\ .
\label{eq36}
\end{equation}
We demonstrate in Appendix~A that the value of $\xi$ can be estimated
using
\begin{equation}
\xi = {1 \over {\cal M_\infty}^2}
\ ,
\label{eq37}
\end{equation}
where ${\cal M_\infty}$ denotes the incident (upstream) Mach number of
the flow with respect to the radiation sound speed. In the sinking
region below the shock, the effective velocity approaches zero as the
gas settles onto the stellar surface. The relationship between the
upstream Mach number ${\cal M_\infty}$ and the X-ray luminosity $\LX$ is
plotted in Fig.~12 from Becker (1998). For the parameter range of
interest here, it is sufficient to adopt a constant value for $\xi$ in
the range $\xi \sim 10^{-2} - 10^{-3}$. The low value for the effective
velocity tends to make the emission region more compact in the supercritical
sources.

Combining relations, we find that
\begin{equation}
\int_0^{\hs} {\pi \mp c \over \dot M \sigmaT \xi}
\, dz' = 1
\ ,
\label{eq38}
\end{equation}
and therefore the altitude of the emission region is given by
\begin{equation}
\hs = {\dot M \sigmaT \xi \over \pi \mp c}
= {\LX R_* \sigmaT \xi \over \pi \mp c G M_*}
\ ,
\label{eq39}
\end{equation}
where the final result follows from Eq.~(\ref{eq15}). We can also
express $\hs$ in cgs units using
\begin{equation}
\begin{aligned}
\hs = 2.28 \times 10^3 \ {\rm cm} \
& \left(\xi \over 0.01\right) \,
\left(M_* \over 1.4 M_\odot\right)^{-1} \\
\times \
& \left(R_* \over 10\,{\rm km}\right)
\left(\LX \over 10^{37} {\rm erg\ sec^{-1}}\right)
\ .
\end{aligned}
\label{eq40}
\end{equation}
Note that the emission height in the supercritical sources varies in
proportion to the luminosity $\LX$ (see Fig.~1).

Based on Eqs.~(\ref{eq16}) and (\ref{eq40}), we conclude that
\begin{equation}
{\hs \over H} = 2.0 \, \xi \ll 1
\ ,
\label{eq41}
\end{equation}
and therefore the characteristic height of emission in the
supercritical sources is located far below the altitude of the
radiation-dominated shock. Eq.~(\ref{eq40}) indicates that the height of
the emission region $\hs$ scales in proportion to the luminosity $\LX$
in the supercritical sources, which is consistent with the observed
behavior in the group 1 sources (Klochkov et al. 2011).

\subsection{Subcritical sources}

In the subcritical sources (luminosity $\LX \lapprox \Lcrit$), the
matter still passes through a radiation-dominated shock, which
accomplishes the initial deceleration, but the pressure of the radiation
is insufficient to bring the matter to rest at the stellar surface
(Basko \& Sunyaev 1976). In this case, the final stopping occurs via
direct Coulomb interactions close to the base of the accretion column
(Burnard et al. 1991). Our goal in this section is to estimate the
characteristic emission height for the subcritical sources, denoted by
$\hc$, which is the altitude at which Coulomb interactions begin to
decelerate the plasma to rest. The emission is expected to be
concentrated in this region because this is essentially the first
opportunity that the radiation inside the column has to diffuse through
the walls. At lower altitudes, the radiation is trapped in the column
due to the increasing density, and therefore the emission through the
walls tapers off as $z \to 0$. Hence we expect that in the subcritical
case, the CRSF energy is imprinted at the altitude $z=\hc$ where the
strong Coulomb deceleration begins.

The Thomson optical depth, $\tau_*$, required to stop the flow via
Coulomb interactions can be estimated in the typical pulsar magnetic
field regime using Eq.~(3.34) from Nelson et al. (1993) to write
\begin{equation}
\tau_* = 51.4 \left(M_* \over 1.4 M_\odot\right)^{2}
\left(R_* \over 10\,{\rm km}\right)^{-2}
{1 \over \ln(2n_{\rm max})}
\ ,
\label{eq42}
\end{equation}
where the maximum excited Landau level, $n_{\rm max}$, is given by
\begin{equation}
n_{\rm max} = {m_e \vff^2 \over 2 \Ecyc}
\ .
\label{eq43}
\end{equation}
A summary of the derivation leading up to Eq.~(\ref{eq42}) is provided
in Appendix~B. Adopting typical X-ray pulsar parameters, we find that
$\tau_* \sim 20$, which is the value utilized in our numerical examples.

We can use Eq.~(\ref{eq42}) to estimate the emission height in the
subcritical sources, $\hc$, as follows. The Thomson depth $\tau$ as
a function of the altitude $z$ measured from the stellar surface is
computed using
\begin{equation}
\tau(z) = \int_0^z {\rho(z') \sigmaT \over \mp} \, dz'
\ ,
\label{eq44}
\end{equation}
where $\rho = n_e m_p$ is the mass density, given by (see Eq.~(\ref{eq12}))
\begin{equation}
\rho(z) = {\dot M \over \pi r_0^2 v(z)}
\ .
\label{eq45}
\end{equation}

Assuming that the gas decelerates uniformly in the Coulomb stopping
region (starting at altitude $\hc$) from the post-shock velocity $\vps$,
we find that the required deceleration is given by $a=\vps^2/(2 \, \hc)$
(cf. Eq.~(\ref{eq6})). The velocity profile $v(z)$ associated with the
constant deceleration $a$ is computed using the standard kinematical
relations
\begin{equation}
a=-{dv\over dt}=v{dv\over dz}={1\over 2}{dv^2\over dz}
\ ,
\label{eq46}
\end{equation}
where the negative sign appears because we have defined $v$ and $a$ to
be positive. Setting $\vps=\vff/7$ (see Eq.~(\ref{eq4})), we obtain for
the deceleration $a=\vff^2/(98 \, \hc)$. Upon integration of
Eq.~(\ref{eq46}), we therefore find that the velocity profile in the
Coulomb stopping region is given by
\begin{equation}
v(z) = {\vff \over 7} \sqrt{z \over \hc}
\ ,
\label{eq47}
\end{equation}
where $\vff$ is evaluated using Eq.~(\ref{eq3}).

Using Eq.~(\ref{eq47}) to substitute for $v(z)$ in Eq.~({\ref{eq45}) and
carrying out the integration in Eq.~(\ref{eq44}), we obtain for the
optical depth profile
\begin{equation}
\tau(z) = {14 \dot M \sigmaT \over \pi r_0^2 \mp}
\left(2GM_* \over R_*\right)^{-1/2}
\sqrt{\hc z}
\ .
\label{eq48}
\end{equation}
Finally, setting $z=\hc$ and $\tau=\tau_*$, we find that the Thomson
optical depth required for Coulomb stopping is given by
\begin{equation}
\tau_* = {14 \dot M \sigmaT \hc \over \pi r_0^2 \mp}
\left(2GM_* \over R_*\right)^{-1/2}
\ ,
\label{eq49}
\end{equation}
which can be rearranged to obtain for the subcritical emission height
\begin{equation}
\hc = {\pi r_0^2 \mp \tau_* \over 14 \dot M \sigmaT}
\left(2GM_* \over R_*\right)^{1/2}
\ .
\label{eq50}
\end{equation}
Substituting for $\dot M$ and $r_0$ using Eqs.~(\ref{eq15}) and
(\ref{eq23}), respectively, yields the equivalent cgs expression
\begin{equation}
\begin{aligned}
\hc = 1.48 \times 10^5 \ {\rm cm} \
& \left(\Lambda \over 0.1
\right)^{-1} \left(\tau_* \over 20\right)
\left(M_* \over 1.4 M_\odot\right)^{19/14}
\left(R_* \over 10\,{\rm km}\right)^{1/14} \\
\times \
& \left(B_* \over 10^{12}{\rm G}\right)^{-4/7}
\left(\LX \over 10^{37} {\rm erg\ sec^{-1}}\right)^{-5/7}
\label{eq51}
\ .
\end{aligned}
\end{equation}
This result indicates that the emission height in the subcritical
sources decreases with increasing luminosity, which is consistent with
the behavior observed in the group 2 sources (Staubert et al. 2007). As
indicated in Fig.~1, in the subcritical case, a decrease in the
luminosity causes the beam pattern to transition from a pure fan
configuration to a hybrid pattern that includes a pencil component. At
very low luminosities, the shock essentially sits on the stellar
surface, and the emission occurs via the pencil component only (Burnard
et al. 1991; Nelson et al. 1993).

\subsection{Transition from Coulomb stopping to gas shock}

Eq.~(\ref{eq51}) gives the height of the emission zone in the
subcritical case, based on the assumption that the final deceleration to
rest at the stellar surface occurs via Coulomb interactions. We can
estimate the minimum luminosity (or accretion rate) such that Coulomb
interactions are capable of stopping the flow by requiring that $\hc <
H$, where $H$ is the altitude of the radiation-dominated shock. If this
condition is violated, then the incident velocity of the gas entering
the Coulomb deceleration region becomes essentially the full free-fall
velocity, $\vff$, rather than the reduced post-shock velocity, $v_{\rm
ps}=\vff/7$. This in turn causes the Thomson depth $\tau$ to fall below
the value $\tau_* \sim 20$ required for the gas to be effectively
stopped via Coulomb interactions. It is not completely clear what
happens in this case, but we expect that the final phase of deceleration
will occur via passage through a gas-mediated shock near the stellar
surface (Langer \& Rappaport 1982).

By combining Eqs.~(\ref{eq14}), (\ref{eq15}) and (\ref{eq50}),
we can show that the condition $\hc < H$ implies that
\begin{equation}
\LX > {2^{1/4} \pi \mp r_0 \over \sigmaT}
\left(\tau_* c \over 7\right)^{1/2} \left(G M_* \over R_*\right)^{5/4}
\ .
\label{eq52}
\end{equation}
Substituting for $r_0$ using Eq.~(\ref{eq23}) and rearranging,
we obtain
\begin{equation}
\LX > \Lcoul
\ ,
\label{eq53}
\end{equation}
where
\begin{equation}
\begin{aligned}
\Lcoul = 1.17 \times 10^{37} \, {\rm erg\ sec^{-1}}
& \left(\Lambda \over 0.1\right)^{-7/12}
\left(\tau_* \over 20\right)^{7/12}
\left(M_* \over 1.4 M_\odot\right)^{11/8} \\
\times \
& \left(R_* \over 10\,{\rm km}\right)^{-13/24}
\left(B_* \over 10^{12}{\rm G}\right)^{-1/3}
\ .
\end{aligned}
\label{eq54}
\end{equation}
For luminosities $\LX \lapprox \Lcoul$, we expect that the
characteristic emission height settles down onto the stellar surface.
For very low luminosities, $\LX \lapprox 10^{34-35}\rm erg\ sec^{-1}$,
the radiation-dominated shock and the Coulomb atmosphere both dissipate,
and the matter strikes the stellar surface after passing through a
gas-mediated shock (Langer \& Rappaport 1982), as indicated in Fig.~1.

\section{Applications}

Our final result for the critical luminosity as a function of the
surface magnetic field strength $B_*$ is given by Eq.~(\ref{eq32}),
which can be rewritten as
\begin{equation}
\begin{aligned}
\Lcrit = 1.28 \times 10^{37} \, {\rm erg\ sec^{-1}}
& \left(\Lambda \over 0.1\right)^{-7/5} w^{-28/15} \\
\times \ \left(M_* \over 1.4 M_\odot\right)^{29/30}
& \left(R_* \over 10\,{\rm km}\right)^{1/10}
\left(E_* \over {\rm 10 \, keV}\right)^{16/15}
\ ,
\end{aligned}
\label{eq55}
\end{equation}
where
\begin{equation}
E_* = 11.58 \ {\rm keV} \left(B_* \over 10^{12}{\rm G}\right)
\label{eq56}
\end{equation}
denotes the surface value of the CRSF energy. This relation is indicated
by the dashed blue line in Fig.~2. Sources to the right of this line
are radiating supercritically, and consequently radiation pressure
accomplishes the deceleration all the way down to the stellar surface.
For sources to the left of this line, the final deceleration occurs via
Coulomb interactions.

\begin{table*}
\caption{Source sample characteristics.}
\centering
\begin{tabular}{c c l c c c}
\hline
\hline
Source     &  Instrument   & Long-term      & Pulse-pulse &$\Delta E$ &Distance\\
           &               & correlation    & correlation &(keV)      &(kpc)\\
\hline
4U 0115+63 & INTEGRAL/RXTE & cont.dependent$^1$ & negative$^6$  &[3-60]  &8.0$^7$ \\
V0332+53   & INTEGRAL/RXTE & negative$^2$       & negative$^6$  &[3-100] &7.5$^8$ \\
Her X-1    & RXTE          & positive$^3$       & positive$^6$  &[5-60]  &6.4$^9$ \\
A0535+26   & INTEGRAL/RXTE   & no$^4$         & positive$^6$  &[3-50]  &2.0$^{10}$ \\
GX-304-1   & RXTE/Suzaku   & positive$^5$       & --            &[3-100] &2.4$^{11}$ \\
\hline
\end{tabular}\\
{\small $^1$M\"uller et al. 2011,
$^2$Tsygankov et al. 2010,
$^3$Staubert et al. 2007,
$^4$Caballero et al. 2007,
$^5$Yamamoto et al. 2011,
$^6$Klochkov et al. 2011,
$^7$Negueruela \& Okazaki 2001,
$^8$Negueruela et al. 1999,
$^9$Reynolds et al. 1997,
$^{10}$Steele et al. 1998,
$^{11}$Parkes et al. 1980}
\tablefoot{Our analysis is based on a combination of
published data (references given) and reprocessed observational data for
a number of cyclotron line sources [column: Source] as observed by
different X-ray observatories [column: Instrument], listed here. The
cyclotron line sources have been observed to show different types of
correlation, or no correlation, of their cyclotron line energies with
changes in the X-ray luminosity.}
\label{tab:sample}
\end{table*}

It is now interesting to compute the critical luminosity for a number of
sources and to compare it with the observed variability of $\Ecyc$ as a
function of $\LX$ for subcritical and supercritical cases. We have
selected for this study the sources for which the behavior of the
cyclotron line energy with luminosity has been studied sufficiently
well, including both the variations on long timescales (days to months)
and short timescales (``pulse-to-pulse''). Our source sample is listed
in Table~1, where we also summarize the reported $\Ecyc$--$\LX$ behavior
(positive/negative correlation or no dependence) for each source. The
two panels in Fig.~2 depict the variability of $\Ecyc$ with luminosity
both on ``pulse-to-pulse'' (left) and longer (right) timescales for all
of the sources considered here. The corresponding references for the
data are given in Table~1. The data for the longer-term variability are
based on studies of the phase-averaged emission conducted either during
outbursts of the transient sources (Tsygankov et al. 2007, 2010), or
following the long-term variation of the emission from the persistent
sources (Staubert et al. 2007). For the pulse-to-pulse variability we
refer to the results of Klochkov et al. (2011) who have shown that for a
set of pulsars the cyclotron line energy varies with the amplitude of
individual pulses. This amplitude most probably reflects the
instantaneous mass accretion rate.

\begin{figure*}
\begin{center}
  \includegraphics[angle=0, width=\columnwidth]{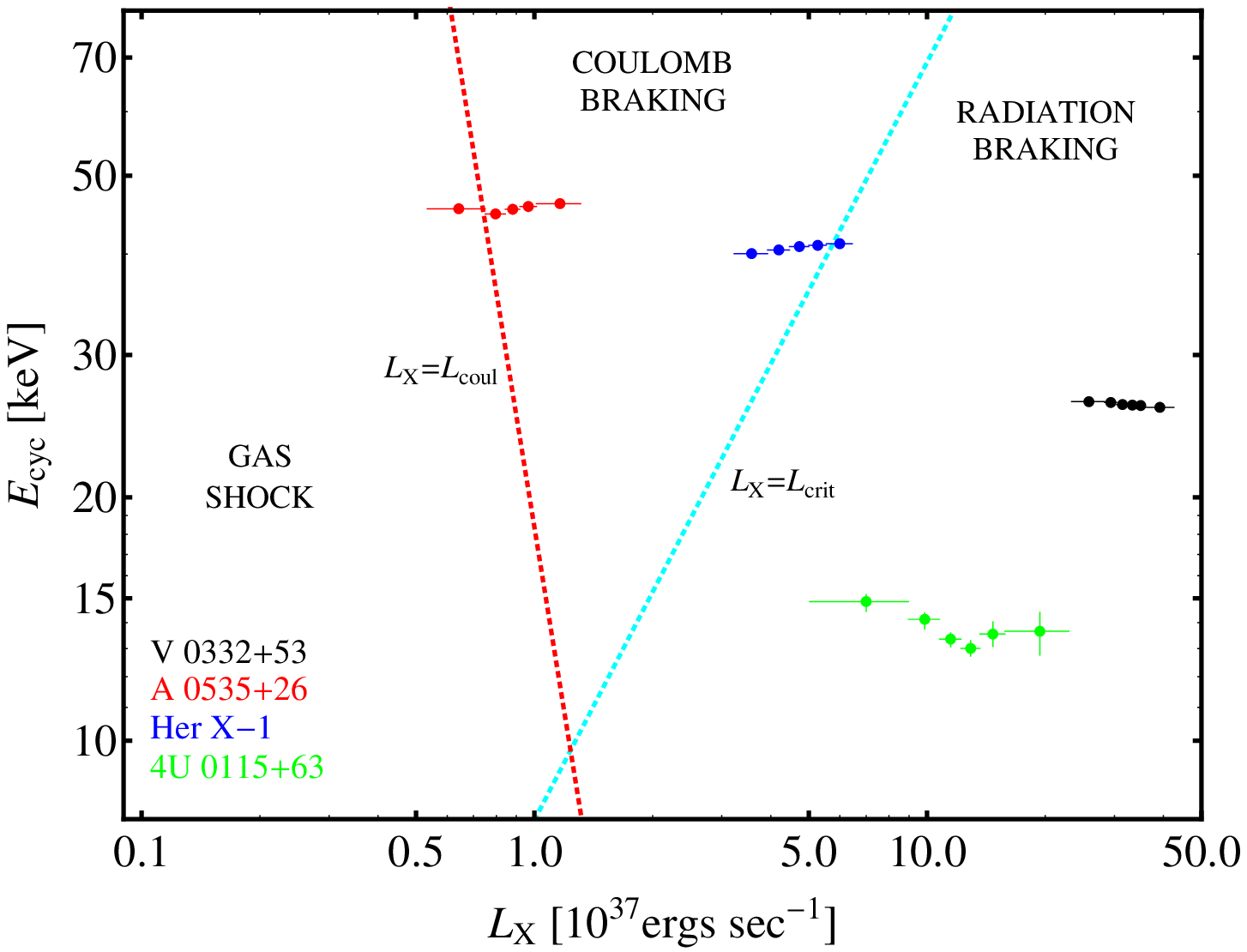}
  \includegraphics[angle=0, width=\columnwidth]{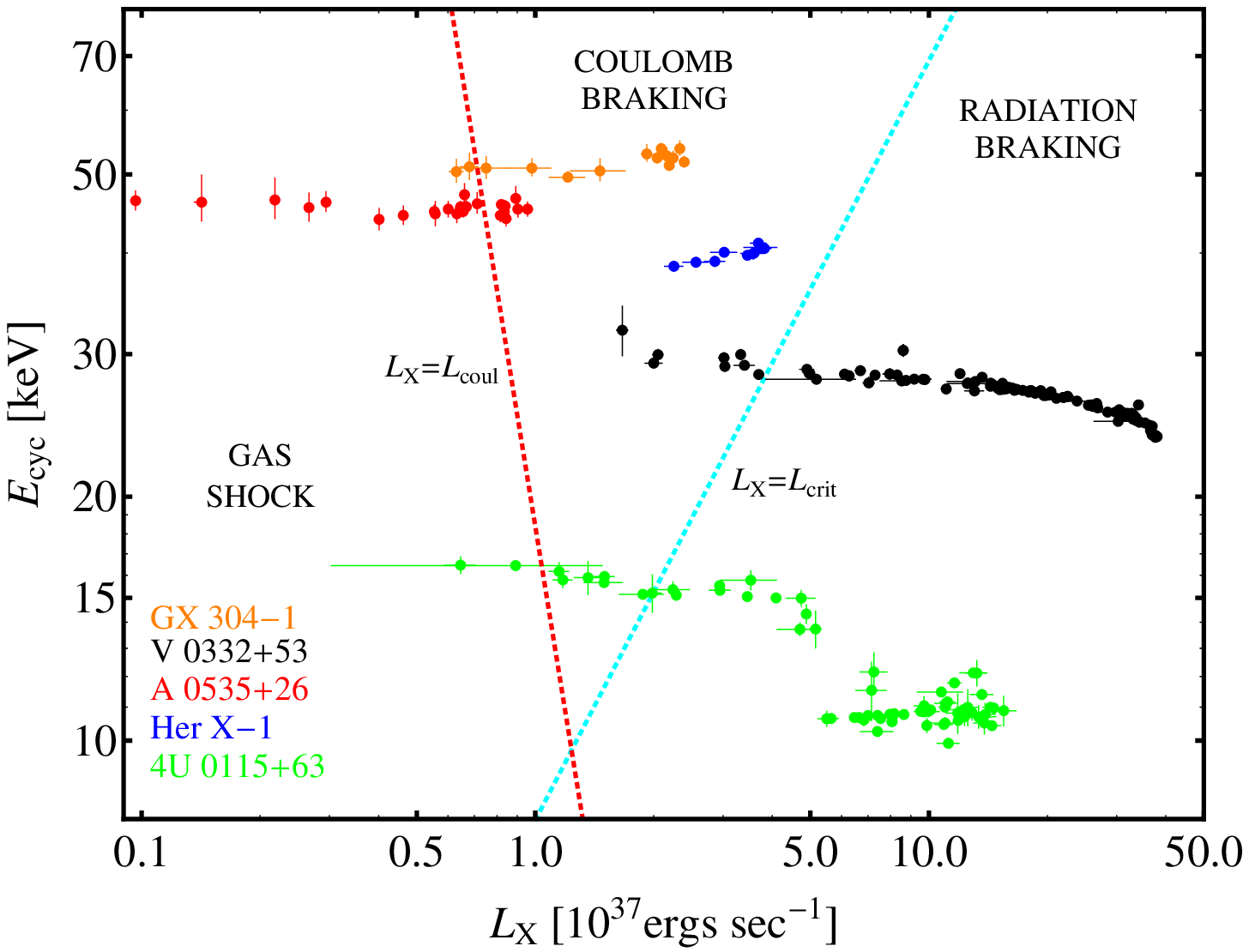}
\caption{Variability of the cyclotron line energy with luminosity for
different sources. The blue dashed line represents the critical
luminosity, plotted by setting $\LX=\Lcrit$ and $\Ecyc=E_*$, where $\Lcrit$
is evaluated using Eq.~(\ref{eq55}). The red dashed line represents the Coulomb
stopping luminosity, plotted by setting $\LX=\Lcrit$ and $\Ecyc=E_*$, where $\Lcoul$
is evaluated using Eq.~(\ref{eq59}). Left: observations on
a pulse-to-pulse timescale. Right: observations on longer timescales.}
\label{fig:old}
\end{center}
\end{figure*}

The luminosities for A\,0535+26, 4\,U0115+63 and Her X-1 were calculated
by integrating the flux of each source over a nearly identical energy
range $\Delta E$ (see Table~1). For V\,0332+53 (Tsygankov et al. 2010)
and GX\,304-1 (Yamamoto et al. 2011) published values for $\LX$ and
$\Ecyc$ were taken, where the luminosity has been calculated over a
slightly larger energy range [3--100]\,keV. The luminosities were
computed using the source distances listed in Table~1. It should be
noted that uncertainties in the distances will create additional
uncertainties in our computed luminosities, which have not been
considered here. However, the typical effective uncertainties of $\sim
10\%$ in the calculated luminosities do not strongly affect our results.

One can see that the sign of the correlation between the cyclotron line
energy and the luminosity, when detected, is consistent between the
long-term and pulse-to-pulse studies. This suggests that both are
reflecting the same underlying physics. It is clear that the variation
of $\Ecyc$ with $\LX$ for sources on each side of the critical line $\LX
= \Lcrit$ is qualitatively consistent with the theoretical predictions,
i.e., the correlation between $\Ecyc$ and $\LX$ is negative for the
supercritical sources, and positive for the subcritical ones, reflecting
the positive variation of the emission height $h=\hs$ (Eq.~(\ref{eq40}))
with increasing $\LX$ for the former sources and the negative variation
of the emission height $h=\hc$ (Eq.~(\ref{eq51})) for the latter
sources. In the next section, we explore the variation of $\Ecyc$ as a
function of $\LX$ in more detail.

\subsection{Variation of $\Ecyc$ with $\LX$}

The observed value of $\Ecyc$ is connected with the local field strength
$B$ at the emission altitude $h$ via Eq.~(\ref{eq26}), where $B$ has
the dipole dependence
\begin{equation}
\begin{aligned}
{B(R) \over B_*} = \left(R \over R_*\right)^{-3} \ , \ \ \ \ \
R = R_* + h
\ .
\end{aligned}
\label{eq57}
\end{equation}
The corresponding variation of $\Ecyc$ as a function of the emission
height $h$ is therefore given by
\begin{equation}
\begin{aligned}
{\Ecyc \over E_*} = \left(R_* + h \over R_*\right)^{-3}
\ ,
\end{aligned}
\label{eq58}
\end{equation}
where $E_*$ is the surface value for the cyclotron energy, computed
using Equation~(\ref{eq56}). Note that we have neglected the variation
of the gravitational redshift factor, which is reasonable given the
level of approximation employed here (Staubert et al. 2007). In applying
Eq.~(\ref{eq58}) to the supercritical and subcritical cases, we set
$h=\hs$ (Eq.~(\ref{eq40})) and $h=\hc$ (Eq.~(\ref{eq51})), respectively.
Hence Eq.~(\ref{eq58}) can be used to develop theoretical predictions for
the variation of $\Ecyc$ as a function of $\LX$ for supercritical and
subcritical sources.

In Fig.~3 we replot the Fig.~2 data based on a rescaling of the vertical
and horizontal axes using the parameters $E_*$ and $\Lcrit$,
respectively, which are related to each other via Eq.~(\ref{eq55}).
The value of $\Lcrit$ used for each source is listed in Table~2, along
with the corresponding value for the surface cyclotron energy $E_*$
obtained by comparing the theoretical variation of $\Ecyc$ with the
observed variation for each source. In computing $\Lcrit$, we assume for
all sources the same canonical neutron star mass and radius values
$M_*=1.4 \, M_\odot$ and $R_*=10\,$km, and we set $\Lambda=0.1$ and
$w=1$ based on the theoretical considerations discussed above. The
vertical dashed line marked $L_{\rm X}=\Lcrit$ separates the sources
into their sub- and supercritical luminosity states. It should be noted
that, as $M_*$ and $R_*$ are also input parameters for $\Lcrit$, the
exact positioning of the source data on the $x$-axis is driven also by
the assumed canonical mass and radius values, which might in fact differ
between the individual sources.

Fig.~3 also includes curves representing the expected theoretical
variation of the CRSF energy $\Ecyc$ as a function of the luminosity
$\LX$, computed using Eq.~(\ref{eq58}) with either $h=\hs$
(Eq.~(\ref{eq40})) for the supercritical sources or $h=\hc$
(Eq.~(\ref{eq51})) for the subcritical ones. We again adopt the
canonical values $M_* = 1.4\,M_\odot$, $R_* = 10\,$km, $\Lambda=0.1$,
and $w=1$, and we set the Coulomb stopping optical depth using
$\tau_*=20$. The values of $E_*$ and $\xi$ are varied for each source so
as to improve the agreement with the data (see Table~2). The parameter
$\xi$ is only relevant for the supercritical sources. The values of
$\xi$ reported in Table~2 for V\,0332+53 and 4U\,0115+63 are in the
range $\xi \sim 10^{-2} -10^{-3}$, as expected for marginally trapped
accretion columns (Becker 1998). Based on the results depicted in
Fig.~3, we conclude that the agreement between the observed variation of
$\Ecyc$ and that predicted by the theoretical models developed here is
reasonably close for both the supercritical and subcritical sources.

A special case is 4U\,0115+63. This source was previously observed to
show an anticorrelation (e.g. Tsygankov et al. 2007 and references
therein). The right panels of our Figs.~2 and 3 (observations on longer
timescales) include the results for $\Ecyc$ and $\LX$ obtained by
Tsygankov et al. (2007). However, recent studies have shown that the
presence of the anticorrelation depends on the choice of the
continuum model (M\"uller et al. 2011).

\begin{figure*}
\begin{center}
  \includegraphics[angle=0, width=\columnwidth]{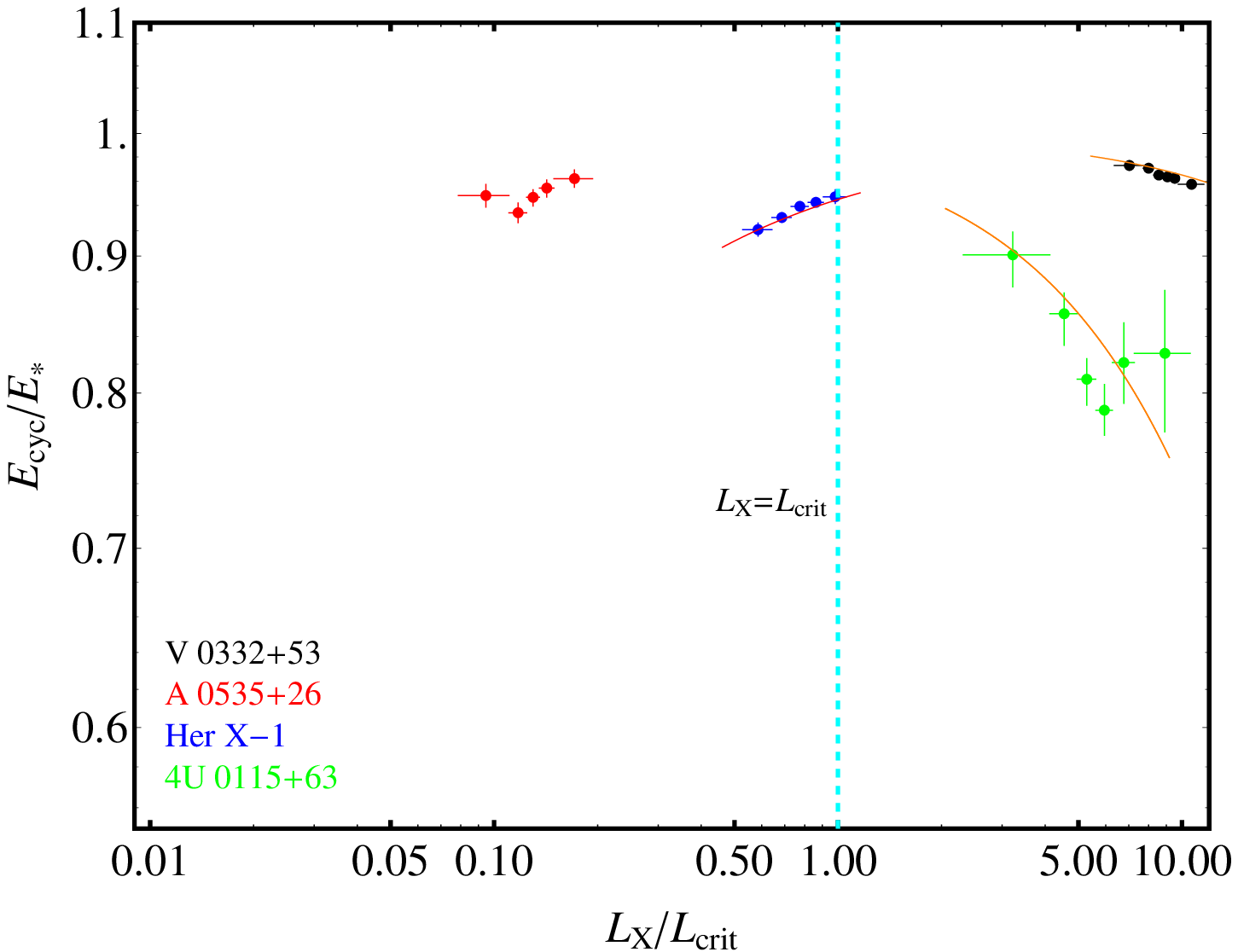}
  \includegraphics[angle=0, width=\columnwidth]{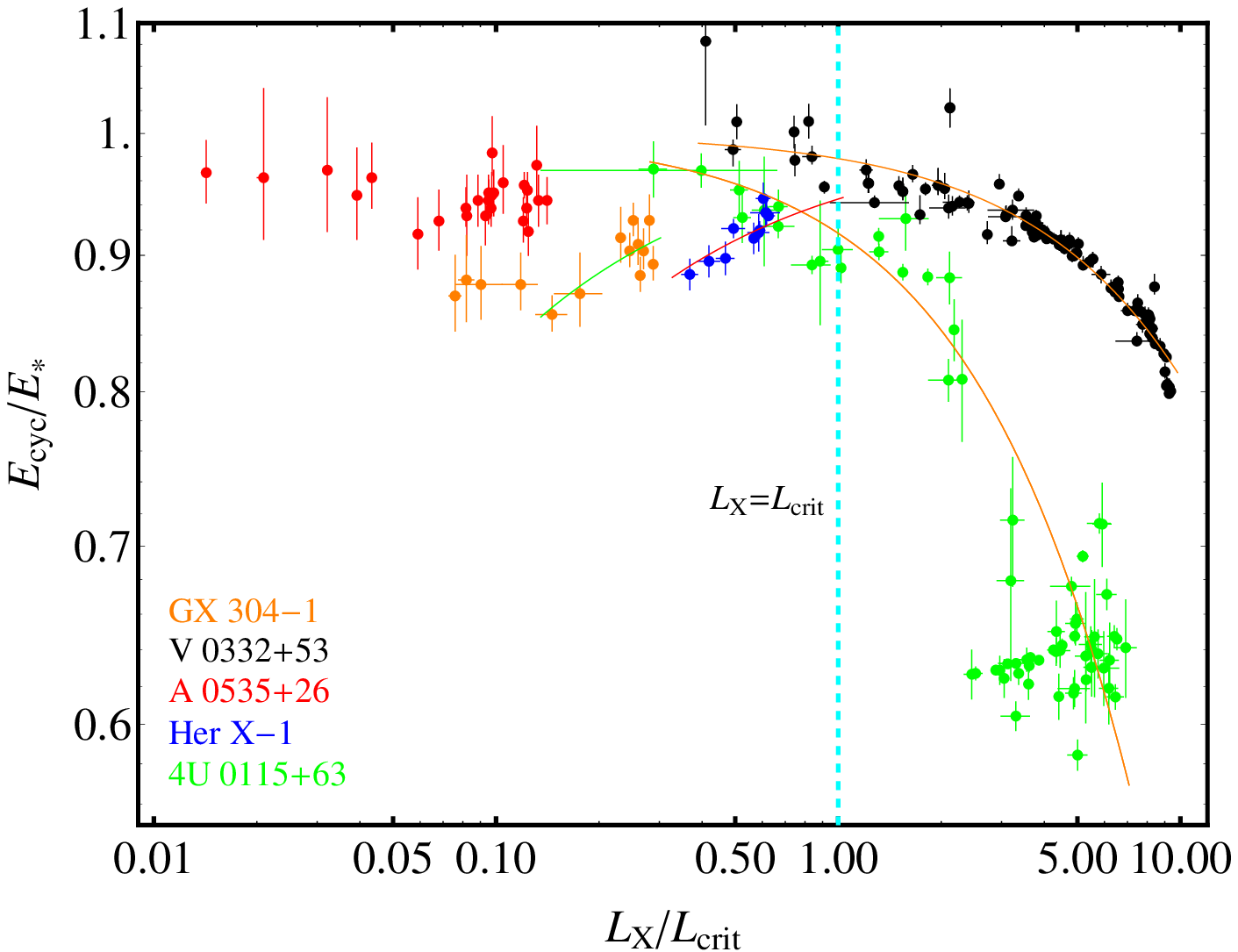}
\caption{Same as Fig.~2 except with the $x$-axis rescaled as
$\LX/\Lcrit$ and the $y$-axis rescaled as $\Ecyc/E_*$.
The values of $E_*$ and $\Lcrit$ used for each source are
listed in Table~2. The solid curves associated with each group of data
represent the theoretical predictions for the variation of the CRSF
energy $\Ecyc$ as a function of $\LX$ for each source,
computed using Eq.~(\ref{eq58}), with the emission height $h$
evaluated using Eq.~(\ref{eq40}) for the supercritical sources
and Eq.~(\ref{eq51}) for the subcritical sources.}
\label{fig:new}
\end{center}
\end{figure*}

\subsection{Comparison of $\LX$ with $\Lcoul$}

It is also interesting to compare the observed luminosities with the
minimum value, $\Lcoul$, required for Coulomb stopping to decelerate the
flow to rest at the stellar surface, given by Eq.~(\ref{eq54}). By
combining Eqs.~(\ref{eq54}) and (\ref{eq56}), we find that $\Lcoul$
is related to $E_*$ via
\begin{equation}
\begin{aligned}
\Lcoul = 1.23 \times 10^{37} \, {\rm erg\ sec^{-1}}
& \left(\Lambda \over 0.1 \right)^{-7/12}
\left(\tau_* \over 20 \right)^{7/12}
\left(M_* \over 1.4 M_\odot\right)^{11/8} \\
\times \
& \left(R_* \over 10\,{\rm km}\right)^{-13/24}
\left(E_* \over {\rm 10 \, keV}\right)^{-1/3}
\ .
\end{aligned}
\label{eq59}
\end{equation}
This relation is indicated by the dashed red line in Fig.~2. For
sources to the left of this line, we expect that the effect of Coulomb
interactions is reduced, and the final stopping occurs via passage
through a discontinuous, gas-mediated shock (Langer \& Rappaport 1982).
Hence we anticipate that the emission region approaches the stellar
surface as $\LX$ is reduced below $\Lcoul$. This interpretation is
consistent with the observational data plotted in Figs.~2 and 3, which
indicate that $\Ecyc$ approaches a constant value in the limit $\LX /
\Lcoul \ll 1$. The left- and right-hand panels in Fig.~2 indicate that
in fact, not much variation of $\Ecyc$ with decreasing luminosity $\LX$
is observed when $\LX \lapprox \Lcoul$, as expected.

Following our hypothesis, sources in their supercritical state ($L_{\rm
X} / \Lcrit \gapprox 1$) should display a negative correlation between
the luminosity and the cyclotron energy while sources in the subcritical
state ($\LX / \Lcrit \lapprox 1$) should display the reverse behavior.
V\,0332+53 and Her X-1 in their super- and subcritical luminosity states
nicely fit into that hypothesis on both long and very short timescales.
In particular, we note that the model parameters listed in Table~2 for
Her X-1 are the same for both the long-term and pulse-to-pulse data.
Hence the model developed here for the variation of $\Ecyc$ as a
function of $\LX$ provides a robust connection with the data across the
entire range of luminosity variation for this source, which implies that
the underlying physical mechanism of variation is the same for the
long-term and pulse-to-pulse variations.

\begin{table*}
\caption{Theoretical parameters for each source, based on analysis of
the pulse-to-pulse variability and the longer-term variability.}
\centering
\begin{tabular}{c c c c c c c}
\hline\hline
Source      & Long-term & Long-term $E_*$  & Long-term $\Lcrit$            & Pulse-pulse & Pulse-pulse $E_*$ &Pulse-pulse $\Lcrit$ \\
            & $\xi$     & [keV]            & [$10^{37}$\,erg\,sec$^{-1}]$  & $\xi$       & [keV]           &[$10^{37}$\,erg\,sec$^{-1}]$   \\
\hline
4U\,0115+63 & $5.72 \times 10^{-2}$       & 17.0        & 2.24        &  $2.14 \times 10^{-2}$  & 16.5  & 2.17  \\
V\,0332+53  & $7.86 \times 10^{-3}$       & 29.7        & 4.06        &  $1.43 \times 10^{-3}$  & 27.0  & 3.67  \\
Her X-1     & -                           & 43.5        & 6.11        &  -                      & 43.5  & 6.11  \\
A\,0535+26  & -                           & 48.0        & 6.78        &  -                      & 48.0  & 6.78  \\
GX\,304-1   & -                           & 58.0        & 8.30        &  -                      & -     & -     \\
\hline
\end{tabular}\\
\tablefoot{The parameter $\xi$ is relevant only for the two
supercritical sources, 4U\,0115+63 and V\,0332+53 (see Sect.~3.1).}
\label{tab:sample2}
\end{table*}

The subcritical source A\,0535+26 shows no significant trend on long
timescales (Fig.~2, right), perhaps due to the fact that the luminosity
is always close to or below the Coulomb stopping limit, $\Lcoul$, which
suggests that we should expect little variation of $\Ecyc$ with $\LX$.
On pulse-to-pulse timescales, A\,0535+26 shows some suggestion of a
positive correlation (Fig.~2, left), as expected for a subcritical
source. The positive correlation suggested by the short-timescale data
may reflect the fact that the luminosity is somewhat higher than
$\Lcoul$, which places it in the subcritical regime according to our
theory. However, we note that our model for the variation of $\Ecyc$
with $\LX$ does not work well for A\,0535+26, unless we choose an
unreasonably large value for $E_*$ relative to the observational data.
We believe this reflects the inapplicability of our model in very
low-luminosity sources with $\LX \lapprox \Lcoul$. For GX\,304-1, only
an indication of a positive $\Ecyc$--$\LX$ correlation, consistent with
its subcritical state, can be seen, as also reported by Yamamoto et al.
2011. No pulse-to-pulse spectra are yet available for GX\,304-1.

\section{Conclusions}

We have examined the hypothesis that observed bimodal variability of the
CRSF energy $\Ecyc$ with luminosity $\LX$ in accretion-powered X-ray
pulsars reflects the dominant mode of accretion, as proposed by Staubert
et al. (2007) and Klochkov et al. (2011). In particular, we have derived
an expression for the critical luminosity $\Lcrit$ such that the
dynamics in the supercritical sources is determined by the radiation
pressure, and the dynamics in the subcritical sources is determined by a
combination of radiation pressure and either Coulomb interactions or gas
pressure. The detailed formula for $\Lcrit$ is given by
Eq.~(\ref{eq32}), but essentially we find that for typical neutron star
parameters, $\Lcrit \sim 1.5 \times 10^{37} B_{12}^{16/15}\,\rm erg\
sec^{-1}$, where $B_{12}$ is the surface magnetic field strength in
units of $10^{12}\,$G.

The formula for the critical luminosity was evaluated for 5 sources,
based on the maximum value for the CRSF centroid energy for each source,
$E_*$, which is treated as a variable parameter in our approach. The
results obtained for $E_*$ are close to the maximum observed values for
the CRSF energy. The results depicted in Fig.~2 confirm that $\LX >
\Lcrit$ in the group 1 sources and $\LX < \Lcrit$ in the group 2
sources. The situation is less clear for highly variable sources with
luminosity $\LX$ that crosses over the line $\LX = L_{\rm crit}$, such
as V\,0332+53 and 4U\,0115+63. These two sources display a negative
correlation between $\Ecyc$ and $\LX$ in the supercritical regime, as
expected, but the trend does not reverse as predicted by our model when
$\LX < \Lcrit$. This suggests that these sources may actually always
remain supercritical, despite the fact that they cross the vertical line
in Fig.~3. This behavior can be accommodated within our model by
slightly changing the parameters $\Lambda$ and $w$ in Eq.~(\ref{eq55}).
Or, alternatively, the behavior of these sources could indicate that
their mass and radius values deviate from the canonical values assumed
here.

We have developed simple physical models describing the quantitative
variation of $\Ecyc$ with $\LX$ in the supercritical and subcritical
sources, given by Equation~(\ref{eq58}), with the emission height $h$
set using $h=\hs$ (Equation~(\ref{eq40})) for the supercritical sources
(group 1), and $h=\hc$ (Equation~(\ref{eq51})) for the subcritical
sources (group 2). In Fig.~3 the formulas we derived for $\Ecyc$ as a
function of $\LX$ were compared with the data for the supercritical and
subcritical sources. The agreement between the theoretical predictions
and the data suggests that our fundamental model for the physical
processes operating in these systems is essentially correct.

Although the observational picture is still not complete, especially for
sources with highly variable luminosities, nonetheless we believe that
the emerging bimodal paradigm for the variability of the CRSF energy
with luminosity in XRBPs supports the hypothesis that we are seeing
direct evidence for two different accretion regimes, depending on
whether the luminosity is above or below the corresponding value of
$\Lcrit$ for the given surface magnetic field strength $B_*$. The
agreement between the theoretical predictions and the observational data
is rather surprising given the level of approximation employed here.
Hence we believe that the ideas explored here can provide a useful
framework for future detailed modeling of these sources, as well as
motivation for further observations of these and other promising
candidates.

\begin{acknowledgements}
The authors gratefully acknowledge generous support provided by the ISSI
in Bern, Switzerland, during the course of this work. MTW acknowledges
support from the US Office of Naval Research. IC acknowledges financial
support from the French Space Agency CNES through CNRS. The authors are
also grateful for assistance and useful comments from S.~M\"uller,
B.~West, K.~Wolfram, and A.~Bodaghee. We also thank S. Tsygankov for
providing observational data in digital form. Finally, we are grateful
to the referee, Lev Titarchuk, for a careful reading of the manuscript
and several insightful suggestions for improvement.

\end{acknowledgements}

\begin{appendix}

\section{Effective Velocity}

The characteristic emission height in the supercritical sources treated
in Sect.~3.1 is estimated by comparing the vertical transport time for
the radiation with the mean escape time for the photons to diffuse out
through the walls of the accretion column. This requires a determination
of the ``effective velocity,'' $\veff$, which is defined as the net
photon transport velocity in the vertical direction, taking into account
the competing effects of advection and diffusion. The former process
tends to drag photons downward toward the stellar surface, and the
latter process tends to transport photons in the opposite direction,
upward through the accretion column. The relationship between the
effective velocity $\veff$ and the flow velocity $v$ is expressed by the
dimensionless parameter $\xi$, defined by
\begin{equation}
\xi \equiv {\veff \over v}
\ .
\label{eqA1}
\end{equation}
In ``trapped'' regions of the flow, vertical advection and diffusion are
nearly balanced, and consequently $\veff \ll v$ and $\xi \ll 1$ (Becker
1998). Trapping tends to occur in the lower, hydrostatic region of the
accretion column in the supercritical sources. In this situation, the
photons tend to ``hover'' in a small altitude range until they escape
through the walls of the accretion column. Hence this effect reduces the
size of the emission region in the supercritical sources treated in
Sect.~3.1.

The gas enters the top of the accretion column moving supersonically,
but it must come to rest at the stellar surface. It follows that the
flow passes through a sonic point somewhere in the column. The sonic
point is located in the middle of the radiation-dominated shock, where
the flow begins to decelerate from the incident free-fall velocity
$\vff$ (Eq.~(\ref{eq3})). Hence the sonic point represents the top of
the hydrostatic sinking region, where the radiation tends to escape, and
we will therefore estimate the value of $\xi$ using conditions there.

In order to determine the flow velocity at the sonic point, it is useful
to consider the conservation of mass and momentum in the hydrostatic
region of the column. We have
\begin{equation}
J \equiv \rho \, v = {\rm const.} \ , \ \ \ \ \ 
I \equiv \Pr + \rho \, v^2 = {\rm const.}
\ ,
\label{eqA2}
\end{equation}
where $J$ and $I$ denote the fluxes of mass and momentum, respectively.
These two fluxes are conserved in the roughly cylindrical, hydrostatic
portion of the accretion column.

We can use Eqs.~(\ref{eqA2}) to obtain a relationship between the flow
velocity $v$ and the radiation Mach number, ${\cal M}$, defined by
\begin{equation}
{\cal M} \equiv {v \over a}
\ ,
\label{eqA3}
\end{equation}
where $a$ denotes the radiation sound speed, given by
\begin{equation}
a = \sqrt{\gamma \Pr \over \rho} \ , \ \ \ \ \
\gamma = {4 \over 3}
\ .
\label{eqA4}
\end{equation}
The result obtained is
\begin{equation}
{I \over J} = v \left(1 + {1 \over \gamma {\cal M}^2}\right)
= {7 \over 4} \, v_c
\ ,
\label{eqA5}
\end{equation}
where $v_c$ denotes the flow velocity at the radiation sonic point,
where ${\cal M}=1$.

The value of $\xi$ can be estimated by examining the vertical
propagating of the photons in a radiation-dominated accretion column
described by the exact dynamical solution obtained by Basko \& Sunyaev
(1976) and Becker (1998). This solution assumes a cylindrical geometry
in the hydrostatic lower region of the accretion column. The total
radiation energy flux in the vertical direction is given by
\begin{equation}
E_r = 4 \Pr v + {c \over n_e \sigmapar} {d\Pr \over dz}
\ ,
\label{eqA6}
\end{equation}
where $\Pr$ is the radiation pressure, and the first and second terms on
the right-hand side represent advection and diffusion, respectively. We
define $\veff$ by writing
\begin{equation}
4 \Pr \veff \equiv E_r
\ ,
\label{eqA7}
\end{equation}
so that $\veff$ represents the ``effective'' bulk velocity that would
yield the correct energy flux.

By combining Eqs.~(\ref{eqA6}) and
(\ref{eqA7}), we obtain
\begin{equation}
\veff  = v + {c \over 4 \, n_e \sigmapar \Pr} {d\Pr \over dz}
\ ,
\label{eqA8}
\end{equation}
or, equivalently,
\begin{equation}
\xi = {\veff \over v} = 1 + {c m_p \over 4 \, \sigmapar J \Pr} {d\Pr \over dz}
\ ,
\label{eqA9}
\end{equation}
where we have eliminated the electron number density using the relation
$J = n_e m_p v$. The pressure $\Pr$ can be expressed in terms of the
flow velocity $v$ by using Eqs.~(\ref{eqA2}) to write
\begin{equation}
\Pr(z) = I - J \, v(z)
\ .
\label{eqA10}
\end{equation}
Using this relation to substitute for the pressure $\Pr$ in
Eq.~(\ref{eqA9}) yields
\begin{equation}
\xi = 1 - {c m_p \over 4 \, \sigmapar (I-Jv)} {dv \over dz}
\ .
\label{eqA11}
\end{equation}

The exact solution for the flow velocity profile $v(z)$ in a cylindrical
accretion column is given by (Basko \& Sunyaev 1976; Becker 1998)
\begin{equation}
v(z) = v_c \, \left({14 \over 7 + 2\epsilon_c}\right)
\, \left[
1-\left({14 \over 7 - 2\epsilon_c}\right)^{-z/z_{\rm st}}
\right]
\ ,
\label{eqA12}
\end{equation}
where
\begin{equation}
\epsilon_c = {3 \, m_p^2 c^2 \over 8 r_0^2 J^2 \sigmaperp \sigmapar}
\label{eqA13}
\end{equation}
denotes the value of the dimensionless total energy flux $(E_r+\rho
v^3/2)/(J v_c^2)$ at the sonic point, and
\begin{equation}
z_{\rm st} = r_0 \left({8 \epsilon_c \sigmaperp \over
3 \sigmapar}\right)^{1/2}\left({2 \over 7 + 2\epsilon_c}\right)
\ln\left({14 \over 7 - 2\epsilon_c}\right)
\label{eqA14}
\end{equation}
is the altitude of the sonic point above the stellar surface.

Combining Eqs.~(\ref{eqA11}), (\ref{eqA12}), (\ref{eqA13}), and
(\ref{eqA14}), we obtain after some algebra
\begin{equation}
\xi = 1 - \left(1 - {4 v \over 7 v_c}\right)^{-1}
\left[1 - \left({7 + 2\epsilon_c \over 14}\right) {v \over v_c}
\right]
\ .
\label{eqA15}
\end{equation}
In particular, at the sonic point ($z=z_{\rm st}$), we have $v=v_c$, and
therefore our result for $\xi$ reduces to
\begin{equation}
\xi \ \bigg|_{z=z_{\rm st}} = {2\epsilon_c - 1 \over 6}
\ .
\label{eqA16}
\end{equation}
The dimensionless energy flux is related to the incident Mach number of
the flow, ${\cal M}_\infty$, via (Becker 1998)
\begin{equation}
\epsilon_c = {1 \over 2} + {3 \over {\cal M}_\infty^2}
\ .
\label{eqA17}
\end{equation}
Combining Eqs.~(\ref{eqA16}) and (\ref{eqA17}) yields for the value of
$\xi$ at the radiation sonic point
\begin{equation}
\xi \ \bigg|_{z=z_{\rm st}} = {1 \over {\cal M_\infty}^2}
\ .
\label{eqA18}
\end{equation}
We use this relation in Sect.~3.1, where we estimate the height of the
emission region in the supercritical sources.

\section{Coulomb Stopping Depth}

In the subcritical sources treated in Sect.~3.2, radiation pressure is
insufficient to decelerate the flow to rest at the stellar surface. In
this regime, the final deceleration likely occurs via Coulomb
interactions between the infalling plasma and the mound of dense gas
that has built up just above the stellar surface. We can estimate the
vertical extent of this region, and therefore obtain an approximation of
the characteristic emission altitude in the subcritical sources, by
computing the Thomson optical depth, $\tau$, measured from the stellar
surface, and setting it equal to the value required to stop the flow,
denoted by $\tau_*$. Nelson et al. (1993) carried out a detailed
calculation of $\tau_*$, and the final result is presented in their
equation Eq.~(3.34). In this section we summarize the derivation.

In a magnetized pulsar accretion column, the Coulomb stopping of the gas
occurs via coupling between infalling protons and stationary electrons
in the mound. The corresponding rate of change of the proton kinetic
energy,
\begin{equation}
E_p = {1 \over 2} \, m_p v^2
\ ,
\label{eqB1}
\end{equation}
is given by Eq.~(3.31) from Nelson et al. (1993), which states that
\begin{equation}
{dE_p \over dz} = {4 \pi n_e e^4 \over m_e v^2} \, \ln\Lambda_c
\ ,
\label{eqB2}
\end{equation}
where $\ln\Lambda_c$ is the Coulomb logarithm and $v$ is the velocity of
the protons. Note that the right-hand side of Eq.~(\ref{eqB2}) is
positive in our sign convention since the value of $E_p$ decreases with
decreasing altitude.

Using Equation~(\ref{eqB1}) to substitute for $v$ yields the equivalent
form
\begin{equation}
{dE_p^2 \over dz} = {m_p \over m_e} \ 4 \pi n_e e^4 \, \ln\Lambda_c
\ .
\label{eqB3}
\end{equation}
We can transform from the altitude $dz$ to the Thomson depth $d\tau$ using
\begin{equation}
d\tau = n_e \sigmaT dz
\ ,
\label{eqB4}
\end{equation}
which yields
\begin{equation}
{dE_p^2 \over d\tau} = {m_p \over m_e} \ {4 \pi e^4 \over
\sigmaT} \, \ln\Lambda_c
\ .
\label{eqB5}
\end{equation}

Treating the Coulomb logarithm as a constant and integrating with respect
to $\tau$, we obtain the solution
\begin{equation}
E_p(\tau) = E_0 \left(1 - {\tau \over \tau_*}\right)^{1/2}
\ ,
\label{eqB6}
\end{equation}
where the stopping depth, $\tau_*$, is defined by
\begin{equation}
\tau_* = {m_e \over m_p} \ {\sigmaT E_0^2 \over
4 \pi e^4 \ln\Lambda_c}
\ ,
\label{eqB7}
\end{equation}
and the incident proton kinetic energy, $E_0$, is equal to the free-fall
value,
\begin{equation}
E_0 = {1 \over 2} \, m_p \vff^2
\ .
\label{eqB8}
\end{equation}
Substituting for the Thomson cross section, $\sigmaT$, in
Eq.~(\ref{eqB7}) using
\begin{equation}
\sigmaT = {8 \pi e^4 \over 3 c^4 m_e^2}
\ ,
\label{eqB9}
\end{equation}
yields the equivalent result
\begin{equation}
\tau_* = {1 \over 6 \ln\Lambda_c} {m_p \over m_e} {\vff^4 \over c^4}
\ .
\label{eqB10}
\end{equation}

In magnetized pulsar accretion columns, with discrete Landau levels, the
Coulomb logarithm is given by Eq.~(3.32) from Nelson et al. (1993),
which states that
\begin{equation}
\ln\Lambda_c = \ln(2 n_{\rm max})
\ ,
\label{eqB11}
\end{equation}
where the maximum excited Landau level, $n_{\rm max}$, is given by
\begin{equation}
n_{\rm max} = {m_e \vff^2 \over 2 \Ecyc}
\ .
\label{eqB12}
\end{equation}
Combining Eqs.~(\ref{eqB10}) and (\ref{eqB11}) and substituting for
$\vff$ using Eq.~(\ref{eq3}) gives the final result,
\begin{equation}
\tau_* = 51.4 \left(M_* \over 1.4 M_\odot\right)^{2}
\left(R_* \over 10\,{\rm km}\right)^{-2}
{1 \over \ln(2n_{\rm max})}
\ ,
\label{eqB13}
\end{equation}
in agreement with Eq.~(3.34) from Nelson et al. (1993). For typical
X-ray pulsar parameters, we obtain $\tau_* \sim 20$, and this is the
value utilized in computing the characteristic emission height in the
subcritical sources in Sect.~3.2.

\end{appendix}

\end{document}